\documentclass[fleqn,usenatbib]{mnras}

\usepackage{newtxtext,newtxmath}
\usepackage[T1]{fontenc}
\usepackage{ae,aecompl}

\usepackage{graphicx}   
\usepackage{amsmath}    

\usepackage{amssymb}    
\usepackage{gensymb}

\usepackage{xfrac}
\usepackage{pdflscape}
\usepackage{CJKutf8} 
\usepackage{arydshln} 
\usepackage[dvipsnames]{xcolor} 

\newcommand{\aref}[1]{\hyperref[#1]{Appendix \ref*{#1}}}

\newcommand{\coone}{\mbox{CO(1-0)}}
\newcommand{\cotwo}{\mbox{CO(2-1)}}
\newcommand{\cothree}{\mbox{CO(3-2)}}

\title[WISDOM: Molecular gas of three megamaser galaxies]{WISDOM project - XVIII.\ Molecular gas distributions and kinematics of three megamaser galaxies}

\author[F.-H.\ Liang et al.]{
  Fu-Heng Liang \begin{CJK*}{UTF8}{gbsn}(梁\,赋珩)\end{CJK*},\thanks{Email: \href{mailto:fuheng.liang@physics.ox.ac.uk}{fuheng.liang@physics.ox.ac.uk}; \href{mailto:ericfuhengliang@gmail.com}{ericfuhengliang@gmail.com}}$^{1}$
  Mark D.\ Smith,$^{1}$
  Martin Bureau,$^{1,2}$
  Feng Gao,$^{3}$
  Timothy A.\ Davis,$^{4}$
  \newauthor{
  Michele Cappellari,$^{1}$
  Jacob S.\ Elford,$^{4}$
  Jenny E.\ Greene,$^{5}$
  Satoru Iguchi,$^{6,7}$
  Federico Lelli,$^{8}$
  Anan Lu,$^{9}$}
  \newauthor{
  Ilaria Ruffa,$^{4}$
  Thomas G.\ Williams$^{1}$
  and Hengyue Zhang$^{1}$}
  \\
  $^{1}$Sub-department of Astrophysics, Department of Physics, University of Oxford, Denys Wilkinson Building, Keble Road, Oxford, OX1~3RH, UK\\
  $^{2}$Yonsei Frontier Lab and Department of Astronomy, Yonsei University, 50 Yonsei-ro, Seodaemun-gu, Seoul 03722, Republic of Korea\\
  $^{3}$Hamburger Sternwarte, Universit\"at Hamburg, Gojenbergsweg 112, D-21029, Hamburg, Germany\\
  $^{4}$Cardiff Hub for Astrophysics Research \& Technology, School of Physics \& Astronomy, Cardiff University, Queens Buildings, Cardiff, CF24~3AA, UK\\
  $^{5}$Department of Astrophysical Sciences, Princeton University, Princeton, NJ 08544, USA\\
  $^{6}$Department of Astronomical Science, SOKENDAI (The Graduate University of Advanced Studies), Mitaka, Tokyo 181-8588, Japan\\
  $^{7}$National Astronomical Observatory of Japan, National Institutes of Natural Sciences, Mitaka, Tokyo 181-8588, Japan\\
  $^{8}$INAF, Arcetri Astrophysical Observatory, Largo Enrico Fermi 5, I-50125 Florence, Italy\\
  $^{9}$Trottier Space Institute and Department of Physics, McGill University, 3600 University Street, Montreal, QC H3A~2T8, Canada\\
}

\date{Accepted 2023 November 23. Received 2023 November 23; in original form 2023 July 20}

\pubyear{2023}

\begin{document}
\label{firstpage}
\pagerange{\pageref{firstpage}--\pageref{lastpage}}
\maketitle

\begin{abstract}
The co-evolution of galaxies and supermassive black holes (SMBHs) underpins our understanding of galaxy evolution, but different methods to measure SMBH masses have only infrequently been cross-checked. We attempt to identify targets to cross-check two of the most accurate methods, megamaser and cold molecular gas dynamics. Three promising galaxies are selected from all those with existing megamaser SMBH mass measurements. We present Atacama Large Millimeter/sub-millimeter Array (ALMA) $^{12}$\cotwo{} and $230$-GHz continuum observations with angular resolutions of $\approx0\farcs5$. Every galaxy has an extended rotating molecular gas disc and $230$-GHz continuum source(s), but all also have irregularities and/or non-axisymmetric features: NGC~1194 is highly inclined and has disturbed and lopsided central $^{12}$\cotwo{} emission; NGC~3393 has a nuclear disc with fairly regular but patchy $^{12}$\cotwo{} emission with little gas near the kinematic major axis, faint emission in the very centre and two brighter structures reminiscent of a nuclear ring and/or spiral; NGC~5765B has a strong bar and very bright $^{12}$\cotwo{} emission concentrated along two bisymmetric offset dust lanes and two bisymmetric nuclear spiral arms. $^{12}$\cotwo{} and $^{12}$\cothree{} observations with the James Clerk Maxwell Telescope are compared with the ALMA observations. Because of the disturbed gas kinematics and the impractically long integration times required for higher angular resolution observations, none of the three galaxies is suitable for a future SMBH mass measurement. Nonetheless, increasing the number of molecular gas observations of megamaser galaxies is valuable, and the ubiquitous disturbances suggest a link between large-scale gas properties and the existence of megamasers.
\end{abstract}




\section{Introduction}

Over the past few decades, tight empirical scaling relations have suggested that supermassive black holes (SMBHs) in galaxy centres co-evolve with their host galaxies across cosmic time (see e.g.\ \citealt{Kormendy+2013ARAA51.511} and \citealt{2021FrASS...8..157D} for reviews). The tightest correlation is between SMBH mass and stellar velocity dispersion ($M_\mathrm{BH}$ -- $\sigma_\star$ relation; e.g.\ \citealt{2000ApJ...539L..13G, 2000ApJ...539L...9F, Beifiori+2012MNRAS419.2497, vdBosch2016ApJ831.134}). Numerous studies have probed potential SMBH and host galaxy self-regulating growth mechanisms through feedback from active galactic nuclei (AGN; see e.g.\ \citealt{2012NewAR..56...93A} for a review). Understanding SMBH properties (mass, growth history, feedback, etc.) is thus critical to our understanding of galaxy evolution.

These scaling relations are however built from no more than $\approx160$ SMBHs with reliable direct (i.e.\ dynamical) mass measurements. Moreover, these come from a variety of measurement methods such as stellar dynamics, ionised-gas dynamics, megamaser (hereafter `maser' for short) dynamics and more recently cold molecular gas dynamics, each with different limitations and potential biases. In particular, most methods are only applicable to certain galaxy types, so that only a few of the measurements have been cross-checked (see \citealt{2013ApJ...770...86W} for an earlier summary). These include: (i) `stars vs.\ ionised gas' in IC~1459 \citep{Cappellari+2002ApJ578.787}, NGC~3379 \citep{2006MNRAS.370..559S}, Centaurus~A \citep{2007ApJ...671.1329N, 2009MNRAS.394..660C}, NGC~3998 \citep{2012ApJ...753...79W}, NGC~4335 \citep{2002AJ....124.2524V} and M81 \citep{Devereux+2003AJ125.1226}; (ii) `stars vs.\ ionised gas vs.\ direct imaging' in M87 \citep{2009ApJ...700.1690G, Gebhardt+2011ApJ729.119, 2013ApJ...770...86W, Jeter+2019ApJ882.82, EHT2019ApJL875.6, 2021ApJ...908..139J, 2022ApJ...935...61B, 2023ApJ...945L..35L, 2024MNRAS.527.2341S}; (iii) `stars vs.\ ionised gas vs.\ reverberation mapping' in NGC~3227 and NGC~4151 \citep{Davies+2006ApJ646.754, 2007ApJ...670..105O, 2008ApJS..174...31H}; (iv) `stars vs.\ masers' in NGC~4258 \citep{2009ApJ...693..946S, 2015MNRAS.450..128D}; (v) `ionised gas vs.\ masers' in NGC~4258 \citep{2007A&A...469..405P}; and (vi) `stars vs.\ proper motions' in the Milky Way (Sagittarius~A*; \citealt{2017MNRAS.466.4040F}). Many of these cross-checking attempts were however affected by disturbed ionised-gas kinematics and/or other issues, and thus were not particularly decisive.

Across these methods, maser dynamics is generally regarded as providing the most accurate and precise SMBH mass measurements, as it measures with high accuracy the gas kinematics close to the SMBHs, yielding a typical mass uncertainty of $\lesssim10\%$ dominated by the galaxy distance \citep[e.g.][]{Herrnstein+2005ApJ629.719}. This method spatially resolves the Keplerian rotation of the parsec-scale discs in which the masers are located, within the spheres of influence (SoIs) of the SMBHs. Maser galaxies are however rare, requiring edge-on masing discs and thus a particular type of nuclear activity mostly present in late-type disc galaxies (i.e.\ a Seyfert~2 AGN), with a narrow range of SMBH masses ($\sim10^7$~M$_\odot$). Nonetheless, maser measurements offer valuable benchmarks for cross-checks (and potentially cross-calibration) with other methods, as stressed in many works \citep[e.g.][]{vdBosch+2016ApJ819.11}. However, maser SMBH mass measurements seem to systematically lie below (i.e.\ at smaller SMBH masses than) the $M_\mathrm{BH}$ -- $\sigma_\star$ relation, even when controlling for morphological type \citep[e.g.][]{Greene+2016ApJL826.32}, and therefore may not follow the general trend defined by all other measurements. This may reveal intrinsic scatter in the $M_\mathrm{BH}$ -- $\sigma_\star$ relation, but it may also arise from systematic effects across the different methods. Cross-checks of individual SMBH measurements is thus imperative.

Due to the high angular resolutions afforded by current mm/sub-mm interferometers, cold molecular gas (particularly CO) dynamics has recently been used to weigh SMBHs. Following the first measurement in NGC~4526 \citep{Davis+2013Nature494.328}, new measurements mostly using the exquisite sensitivity and angular resolution of
the Atacama Large Millmeter/sub-millimeter Array (ALMA) have been made by the millimetre-Wave Interferometric Survey of Dark Object Masses (WISDOM; \citealt{Davis+2017MNRAS468.4675, Davis+2018MNRAS473.3818, 2020MNRAS.496.4061D, Onishi+2017MNRAS468.4663, Smith+2019MNRAS485.4359, Smith+2020MNRASsubmitted2, North+2019MNRAS490.319, 2022MNRAS.516.4066L, Ruffa_accepted}) and others \citep[e.g.][]{Onishi+2015ApJ806.39, Barth+2016ApJ.822L.28, Boizelle+2019ApJ...881...10B, Boizelle+2021ApJ...908...19B,  Nagai+2019ApJ883.193, Ruffa+2019MNRAS.489.3739R,  Nguyen+2020ApJ...892...68N, Nguyen+2021MNRAS.504.4123N, Nguyen+2022MNRAS.509.2920N, Cohn+2021ApJ...919...77C, Kabasares+2022ApJ...934..162K}. 
Many of these observations probe CO emission on the same spatial scales as those probed by masers, and the latest in fact does better (Zhang et al.\ in prep.). This method has only weak biases toward particular galaxy types and is conceptually very simple, mainly constrained by the size of each SMBH SoI and the existence of a central regularly rotating molecular gas disc.
Given its increasing popularity, cross-checking CO and maser (as well as other methods) SMBH dynamical measurements is highly desirable.

This paper thus aims to identify promising targets for future CO SMBH mass measurements (utilising higher resolution follow-up observations), from galaxies with existing maser measurements. Simultaneously, this paper reveals the molecular gas properties of several maser host galaxies at $\sim100$~pc scale, essential to probe the cold interstellar medium (ISM) conditions required for masing. In \autoref{sec_targets}, we present a compilation of existing maser SMBH mass measurements and the three targets selected here for further study. In \autoref{sec_observations}, we describe new intermediate-resolution ALMA as well as James Clerk Maxwell Telescope (JCMT) observations of the molecular gas and mm-continuum emission of those three galaxies, along with standard data products. The potential for SMBH
mass measurements using CO observations at higher angular resolutions is discussed in \autoref{sec_discussion}. We discuss the link between molecular gas disc properties and maser emission with an enlarged sample in \autoref{sec:maser_disturb}.
Finally, we summarise and conclude in \autoref{sec_conclusions}.

\section{Targets}
\label{sec_targets}

\subsection{Candidate selection}
\label{sec:candidate}

To cross-check cold molecular gas and maser dynamical SMBH mass measurements, we must first identify galaxies with existing maser measurements that also appear promising for molecular gas measurements. We use the compilation of SMBH mass measurements of \citet{vdBosch2016ApJ831.134} as our starting point, including all maser measurements in their Tables~2 and 3. We update the galaxy distances when better determinations are now available, including those from maser monitoring programmes (NGC~6264, NGC~6323, NGC~5765B and UGC~3789), part of the Megamaser Cosmology Project (MCP; see \citealt{2009astro2010S..23B} and \autoref{tab:maserSOI}), and adjust the SMBH masses accordingly ($M_{\rm BH}\propto D$ for all dynamical mass measurements, where $D$ is the galaxy distance). We remove NGC~1386, as there is no refereed source for its SMBH mass and its $^{12}$\coone{}\footnote{Hereafter we omit the carbon atomic mass number and refer to the $^{12}$CO isotope simply as CO.} has already been imaged with ALMA by \citet{Ramakrishnan+2019MNRAS487.444} and \citet{Zabel+2019MNRAS483.2251}. We also correct the SMBH mass of IC~2560 quoted in \citeauthor{vdBosch2016ApJ831.134} (\citeyear{vdBosch2016ApJ831.134}; $\log(M_\mathrm{BH}/\mathrm{M}_\odot)=7.64\pm0.05$) back to the original mass reported by \citeauthor{Yamauchi+2012PASJ64.103} (\citeyear{Yamauchi+2012PASJ64.103}; $\log(M_\mathrm{BH}/\mathrm{M}_\odot)=6.54\pm0.06$). This megamaser parent sample is summarised in \autoref{tab:maserSOI}.

\begin{table*}
\caption{Predicted SMBH SoIs and selection criteria of maser galaxies.}
\label{tab:maserSOI}
\begin{tabular}{lccccccccl}
\hline
\hline
Galaxy      & Distance       & $\log(M_\mathrm{BH}/{\mathrm M}_\odot)$ & $\sigma_\star$    & $R_\mathrm{SoI}$          & $\theta_\mathrm{SoI}$      & SoI       & Dust       & Dec.        & References \\
            & (Mpc)          &                   & (km s$^{-1}$) & (pc)           & (mas)          &            &            &            &                                  \\
(1)         & (2)            & (3)               & (4)           & (5)            & (6)            & (7)        & (8)        & (9)        & (10)                             \\
\hline
Circinus    & $\phantom{11}2.8\phantom{1}\pm\phantom{1}0.5\phantom{1}$    & $6.06\pm0.07$     & $158\pm18$    & $\phantom{1}0.20\pm0.06$  & $\phantom{1}15\phantom{.1}\pm\phantom{1}5\phantom{.1}$ & \checkmark & \checkmark & \checkmark & (1)  \\
ESO 558-009 & $108\phantom{.11}\pm\phantom{1}6\phantom{.11}$ & $7.23\pm0.03$     & $170\pm20$    & $\phantom{1}2.5\phantom{1}\pm0.6\phantom{1}$  & $\phantom{11}4.9\pm\phantom{1}1.2$    & -          & \checkmark & \checkmark & (2)         \\
IC 1481     & $\phantom{1}79\phantom{.11}\pm\phantom{1}6\phantom{.11}$   & $7.11\pm0.13$     & $\phantom{1}95\pm27$     & $\phantom{1}6.1\phantom{1}\pm3.9\phantom{1}$  & $\phantom{1}16\phantom{.1}\pm10\phantom{.1}$   & \checkmark & -          & \checkmark & (3)   \\
IC 2560     & $\phantom{1}31\phantom{.11}\pm13\phantom{.11}$      & $6.54\pm0.06$     & $141\pm10$    & $\phantom{1}0.90\pm0.18$  & $\phantom{11}6.0\pm\phantom{1}2.8$    & -        & -          & \checkmark & (4)   \\
J0437+2456  & $\phantom{1}65\phantom{.11}\pm\phantom{1}4\phantom{.11}$   & $6.46\pm0.05$     & $110\pm13$    & $\phantom{1}1.03\pm0.27$  & $\phantom{11}3.3\pm\phantom{1}0.9$    & -          & -          & -          & (2)         \\
Mrk1029     & $121\phantom{.11}\pm\phantom{1}7\phantom{.11}$  & $6.28\pm0.12$     & $132\pm15$    & $\phantom{1}0.47\pm0.17$  & $\phantom{11}0.8\pm\phantom{1}0.3$    & -          & -          & \checkmark & (2)         \\
NGC 1068    & $\phantom{1}16\phantom{.11}\pm\phantom{1}9\phantom{.11}$   & $6.95\pm0.02$     & $151\pm\phantom{1}7$     & $\phantom{1}1.67\pm0.17$  & $\phantom{1}22\phantom{.1}\pm13\phantom{.1}$  & \checkmark & -          & \checkmark &  (5)    \\
NGC 1194    & $\phantom{1}58\phantom{.11}\pm\phantom{1}6\phantom{.11}$   & $7.85\pm0.02$     & $148\pm24$    & $14\phantom{.11}\pm5\phantom{.11}$ & $\phantom{1}50\phantom{.1}\pm17\phantom{.1}$  & \checkmark & \checkmark & \checkmark & (6)        \\
NGC 1320    & $\phantom{1}34.2\phantom{1}\pm\phantom{1}1.9\phantom{1}$   & $6.74\pm0.21$     & $141\pm16$    & $\phantom{1}1.2\phantom{1}\pm0.7\phantom{1}$  & $\phantom{11}7\phantom{.1}\pm\phantom{1}4\phantom{.1}$    & -          & -          & \checkmark & (2)         \\
NGC 2273    & $\phantom{1}29.5\phantom{1}\pm\phantom{1}1.9\phantom{1}$   & $6.93\pm0.02$     & $145\pm17$    & $\phantom{1}1.8\phantom{1}\pm0.4\phantom{1}$  & $\phantom{1}12.3\pm\phantom{1}3.0$   & \checkmark & \checkmark & -          & (6)         \\
NGC 2960    & $\phantom{1}67\phantom{.11}\pm\phantom{1}7\phantom{.11}$   & $7.03\pm0.02$     & $151\pm\phantom{1}7$     & $\phantom{1}2.01\pm0.21$  & $\phantom{11}6.2\pm\phantom{1}0.9$    & -          & \checkmark & \checkmark & (6)        \\
NGC 3079    & $\phantom{1}15.9\phantom{1}\pm\phantom{1}1.2\phantom{1}$   & $6.36\pm0.09$     & $145\pm\phantom{1}7$     & $\phantom{1}0.47\pm0.11$  & $\phantom{11}6.1\pm\phantom{1}1.5$    & -          & \checkmark & -          & (7)   \\
NGC 3393    & $\phantom{1}49\phantom{.11}\pm\phantom{1}8\phantom{.11}$   & $7.3\phantom{1}\pm0.4\phantom{1}$     & $148\pm10$    & $\phantom{1}3.6\phantom{1}\pm2.5\phantom{1}$  & $\phantom{1}15\phantom{.1}\pm11\phantom{.1}$  & \checkmark & \checkmark & \checkmark & (3), (8)   \\
NGC 4258    & $\phantom{11}7.3\phantom{1}\pm\phantom{1}0.5\phantom{1}$    & $7.58\pm0.03$     & $115\pm11$    & $12.4\phantom{1}\pm2.4\phantom{1}$ & $351\phantom{.1}\pm73\phantom{.1}$ & \checkmark & -          & -          & (9), (10) \\
NGC 4388    & $\phantom{1}16.5\phantom{1}\pm\phantom{1}1.6\phantom{1}$   & $6.86\pm0.01$     & $107\pm\phantom{1}7$     & $\phantom{1}2.7\phantom{1}\pm0.4\phantom{1}$  & $\phantom{1}34\phantom{.1}\pm\phantom{1}6\phantom{.1}$   & \checkmark & -          & \checkmark & (6)         \\
NGC 4945    & $\phantom{11}3.58\pm\phantom{1}0.22$    & $6.13\pm0.18$     & $135\pm\phantom{1}6$     & $\phantom{1}0.32\pm0.14$  & $\phantom{1}18\phantom{.1}\pm\phantom{1}8\phantom{.1}$   & \checkmark & -          & \checkmark & (10), (11) \\
NGC 5495    & $\phantom{1}96\phantom{.11}\pm\phantom{1}5\phantom{.11}$ & $7.04\pm0.08$     & $166\pm19$    & $\phantom{1}1.7\phantom{1}\pm0.5\phantom{1}$  & $\phantom{11}3.7\pm\phantom{1}1.1$    & -          & -          & \checkmark & (2)         \\
NGC 5765B   & $112\phantom{.11}\pm\phantom{1}5\phantom{.11}$ & $7.61\pm0.04$     & $158\pm18$    & $\phantom{1}7.0\phantom{1}\pm1.7\phantom{1}$ & $\phantom{1}13\phantom{.1}\pm\phantom{1}3\phantom{.1}$   & \checkmark & \checkmark & \checkmark &   (12)      \\
NGC 6264    & $144\phantom{.11}\pm19\phantom{.11}$     & $7.49\pm0.06$     & $158\pm15$    & $\phantom{1}5.3\phantom{1}\pm1.2\phantom{1}$  & $\phantom{11}7.6\pm\phantom{1}2.0$    & -          & \checkmark & -          & (13)         \\
NGC 6323    & $107\phantom{.11}\pm36\phantom{.11}$ & $6.97\pm0.14$       & $158\pm26$    & $\phantom{1}1.6\phantom{1}\pm0.7\phantom{1}$  & $\phantom{11}3.1\pm\phantom{1}1.7$    & -          & -          & -          & (14)         \\
UGC 3789    & $\phantom{1}50\phantom{.11}\pm\phantom{1}5\phantom{.11}$   & $7.06\pm0.05$     & $107\pm12$    & $\phantom{1}4.3\phantom{1}\pm1.1\phantom{1}$  & $\phantom{1}18\phantom{.1}\pm\phantom{1}5\phantom{.1}$   & \checkmark & -          & -          & (15)         \\
UGC 6093    & $152\phantom{.11}\pm15\phantom{.11}$ & $7.41\pm0.03$     & $155\pm18$    & $\phantom{1}4.6\phantom{1}\pm1.1\phantom{1}$  & $\phantom{11}6.3\pm\phantom{1}1.6$    & -          & -          & \checkmark & (16)       \\
\hline \hline
\end{tabular}
  \parbox{\textwidth}{\small Notes: Column~1: galaxy name. Column~2: distance. Column~3: maser-derived SMBH mass. Column~4: stellar velocity dispersion measured within one effective radius, using a variety of methods \citep{vdBosch2016ApJ831.134}. Columns~5 and 6: SMBH SoI physical radius and angular radius. Columns~7 -- 9: selection criterion fulfillment. Column~10: maser SMBH mass measurement references: (1) \citealt{Greenhill+2003ApJ590.162}, (2) \citealt{Gao+2017ApJ834.52}, (3) \citealt{2011AandA...530A.145H}, (4) \citealt{Yamauchi+2012PASJ64.103}, (5) \citealt{2003AandA...398..517L}, (6) \citealt{Kuo+2011ApJ727.20}, (7) \citealt{Yamauchi+2004PASJ56.605}, (8) \citealt{Kondratko+2008ApJ678.87}, (9) \citealt{Herrnstein+2005ApJ629.719}, (10) \citealt{Kormendy+2013ARAA51.511}, (11) \citealt{Greenhill+1997ApJL481.23}, (12) \citealt{Gao+2016ApJ817.128}, (13) \citealt{2013ApJ...767..155K}, (14) \citealt{2015ApJ...800...26K}, (15) \citealt{2013ApJ...767..154R} and (16) \citealt{Zhao+2018ApJ854.124}. We note that the maser emission in some galaxies may be dominated or contaminated by non-disc maser sources, such as masers in outflows/jets; see \citet{2015ApJ...810...65P} for a clean subset with pure disc-maser galaxies and Keplerian rotation.
    }
\end{table*}

We then apply the following selection criteria to retain the best cold molecular gas measurement candidates. Whether each galaxy fulfils each criterion is listed in \autoref{tab:maserSOI}.

\begin{enumerate}
   \item SMBH SoI angular radius $\theta_{\rm SoI} \equiv R_{\rm SoI} / D >0\farcs01$, where the SMBH SoI physical radius $R_{\rm SoI} \equiv{\rm G}M_\mathrm{BH}/\sigma_\star^2$,
   so that the SoI can be spatially resolved using ALMA's longest baselines at \cotwo{} (band~6). 
   \item {\it Hubble Space Telescope} ({\it HST}) imaging (available for all galaxies) showing a regular central dust disc, suggesting a central molecular gas disc in ordered rotation.
   \item Declination $-66\degr<\delta<+20\degr$, to ensure a fairly round ALMA synthesised beam and minimise shadowing.\footnote{We note that even if we relax this criterion to $-90\degr \lid \delta \lid +47\degr$, to reach the absolute declination limit of ALMA, the final sample of galaxies meeting all of our selection criteria does not change.}
\end{enumerate}

The only galaxies to satisfy all these requirements are Circinus, NGC~1194, NGC~3393 and NGC~5765B. Cold molecular gas in Circinus has already been observed with ALMA at high angular resolution ($\approx0\farcs2$) and it is reported to have a disturbed velocity field \citep{2018ApJ...867...48I, 2022A&A...664A.142T}. This galaxy is therefore not suited to cold gas dynamical modelling to derive a SMBH mass, and it is not considered further in this paper. The basic properties of the other three galaxies, for which we obtained and present new ALMA data here, are listed in \autoref{tab:targets}. We also discuss each galaxy in more details below.

\begin{table*}
  \centering
  \caption{Properties of our target galaxies.}
  \label{tab:targets}
  \begin{tabular}{llllll}
    \hline \hline
    Galaxy & Right ascension & Declination & $z_{\rm helio}^1$ & Hubble type & Nuclear activity \\
           & (J2000) & (J2000) &  & \\ \hline
    NGC~1194 & $03^\mathrm{h}03^\mathrm{m}49\fs10870^2$ & $-01\degr06^\prime13\farcs4720^2$ & 0.01363 & S0-$^3$ & Sy2$^4$ \\
    NGC~3393 & $10^\mathrm{h}48^\mathrm{m}23\fs4659^5$ & $-25\degr09^\prime43\farcs477^5$ & 0.01251 & SBa$^{6}$ & Sy2$^{4,7}$ \\
    NGC~5765B & $14^\mathrm{h}50^\mathrm{m}51\fs51884^8$ & $+05\degr06^\prime52\farcs2501^8$ & 0.02754 & Sab$^9$ & Sy2$^{10}$ \\
    \hline \hline
  \end{tabular}
  \parbox{\textwidth}{Notes: 
  (1) Heliocentric redshifts are taken from \citet{2017ApJS..233...25A} for NGC~1194 and NGC~5765B and from \citet{2015ApJ...810...65P} for NGC~3393. 
  (2) Average position of the masers at the systemic velocity of the galaxy (see Tables~1 and 3 of \citealt{Kuo+2011ApJ727.20}).
  (3) \citet{2010ApJS..186..427N}.
  (4) \citet{2018ApJS..235....4O}.
  (5) Average position of the masers at the systemic velocity of the galaxy (see Section~4 of \citealt{Kondratko+2008ApJ678.87}).
  (6) \citet{1991rc3..book.....D}.
  (7) \citet{Baumgartner+2013ApJS207.19}.
  (8) Best-fitting dynamical centre (see Table~7 of \citealt{Gao+2016ApJ817.128}).
  (9) \citet{Pjanka+2017ApJ844.165}.
  (10) \citet{Toba+2014ApJ788.45}.}
\end{table*}

\subsection{NGC~1194}

NGC~1194 is a lenticular galaxy harbouring a Seyfert~2 AGN \citep{2018ApJS..235....4O}, for which we adopt a distance $D=58\pm6$~Mpc. This distance was estimated from NGC~1194's Local Group-centric redshift \citep{1996AJ....111..794K} by \citet{Kormendy+2013ARAA51.511} and \citet{Saglia+2016ApJ818.47} assuming a cosmology derived from the \textit{Wilkinson Microwave Anisotropy Probe} (\textit{WMAP}) 5-year data \citep{2009ApJS..180..330K}. NGC~1194 was first reported to harbour H$_2$O megamasers by \citet{2008ApJ...686L..13G} and is part of the MCP. It has a relatively large maser disc with an inner radius of $0.51$~pc and an outer radius of $1.33$~pc. It hosts the most massive SMBH derived using the maser method to date, $M_\mathrm{BH}=(7.1\pm0.3)\times10^7$~M$_\odot$ at our adopted distance above \citep{Kuo+2011ApJ727.20}. The maser disc has an inclination
$i\approx85\degr$ with a kinematic position angle\footnote{The kinematic (morphological) position angle is measured from north through east until the largest receding velocity (photometric major axis) is reached.} $PA_\mathrm{kin}=337\degr$, while the galaxy's overall inclination is $\approx50\degr$ with a morphological position angle $PA_\mathrm{mor}=145\degr$, as determined from an $r$-band image from the Sloan Digital Sky Survey (SDSS; \citealt{2008ApJS..175..297A}). No warp is detected in the maser disc.

Previous studies have reported atomic hydrogen out to a galactocentric radius of $\approx120\arcsec$ \citep[$\approx4$~kpc;][]{Sun+2013ApJ778.47} and patchy warm molecular hydrogen on a scale of $1\farcs6$ ($450$~pc), limited by the telescope field of view (FoV; \citealt{Greene+2014ApJ788.145}). Ionised gas has been detected through $K$-band emission lines (e.g.\ Br$\gamma$, [\ion{Si}{vi}] and [\ion{Ca}{viii}]) at the galaxy centre by \citet{Greene+2014ApJ788.145} and through the [\ion{O}{iii}] optical emission line over a slightly more extended region ($700\times470$~pc$^2$; \citealt{2003ApJS..148..327S}). \citet{2021ApJS..252...29K} did not detect cold molecular gas using the Atacama Pathfinder Experiment, with a \cotwo{} $3\sigma$ upper limit of $2\times10^7$~K~km~s$^{-1}$~pc$^2$ ($9$~Jy~km~s$^{-1}$), consistent with (i.e.\ larger than) our detected flux reported in \autoref{sec:1194_co}.

\subsection{NGC~3393}

NGC~3393 is an SBa galaxy \citep{1991rc3..book.....D} at an adopted distance $D=49\pm8$~Mpc. This distance was again estimated from the Local Group-centric redshift \citep{1996AJ....111..794K} by \citet{Kormendy+2013ARAA51.511} and \citet{Saglia+2016ApJ818.47} assuming the cosmology derived from \textit{WMAP} 5-year data \citep{2009ApJS..180..330K}.
NGC~3393 has a large-scale stellar bar (\mbox{$PA_\mathrm{mor}\approx160\degr$}), extended radio jets \citep[e.g.][]{Cooke+2000ApJS129.517} and a Seyfert~2 nucleus \citep{Baumgartner+2013ApJS207.19}. A nuclear bar ($PA_\mathrm{mor}\approx145\degr$) has also been posited \citep[e.g.][]{Lasker+2016ApJ825.3}. The presence of two compact X-ray sources separated by $\approx130$~pc suggests there are two SMBHs in the nuclear region \citep{Fabbiano+2011Nature477.431}, but these may be due to noise \citep{Koss+2015ApJ807.149} and subsequent radio, near-infrared, optical, UV and hard X-ray observations are all consistent with a single point source \citep{2014ApJ...780..106I, Koss+2015ApJ807.149}.

H$_2$O megamasers were discovered using the National Aeronautics and Space Administration (NASA) Deep Space Network \citep{Kondratko+2006ApJ638.100}, and mapped with very long baseline interferometry (VLBI) to infer a central SMBH mass $M_\mathrm{BH}=(3.0\pm0.2)\times10^7$~M$_\odot$ at our adopted distance above \citep{Kondratko+2008ApJ678.87}. With the same dataset, \citet{2011AandA...530A.145H} inferred a different SMBH mass of $M_\mathrm{BH}=0.58\times10^7$~M$_\odot$ (at our adopted distance above) using different dynamical modelling. We therefore adopt the mean of these two measurements for the NGC~3393 SMBH mass, and half the difference as the uncertainty, as done by \citet{Kormendy+2013ARAA51.511}. The maser disc is large, with an outer radius of $1.5$~pc, and was assumed to be edge-on with a tentative warp and $PA_\mathrm{kin}\approx326\degr$, perpendicular to both the kiloparsec-scale radio jet and the axis of the narrow-line region (see \citealt{Kondratko+2008ApJ678.87} and references therein). The overall inclination of the galaxy is $i=44\degr$ with \mbox{$PA_\mathrm{mor}\approx160\degr$}, as determined from a Two Micron All Sky Survey (2MASS) $K_\mathrm{s}$-band image \citep{2006AJ....131.1163S}.

\cotwo{} emission was recently mapped with ALMA by \cite{Finlez+2018MNRAS479.3892}. They presented two sets of maps with different imaging parameters, one with a $0\farcs56$ synthesised beam\footnote{Beam sizes quoted in this paper are all full-widths at half-maxima, FWHM.} and $2.5$~km~s$^{-1}$ channels, the other with a $0\farcs68$ synthesised beam and $10$~km~s$^{-1}$ channels. Very little \cotwo{} emission is detected in the very centre,
which they attribute to either molecular gas destruction by the jet or high molecular gas densities/temperatures not detected in the $J=2-1$ CO transition \citep[see e.g.][]{2018MNRAS.476...80M, 2018MNRAS.479.5544M, 2022MNRAS.510.4485R}. Our new observations slightly improve the angular resolution and sensitivity of these observations (see \autoref{tab:observationProperties}). \cite{Finlez+2018MNRAS479.3892} also provided a sophisticated analysis of the ionised-gas kinematics, exploiting abundant optical emission lines. In addition to a regularly rotating component, there are a jet-driven outflow along the jet axis and an equatorial outflow perpendicular to it. \citet{2022MNRAS.510.1716R} reported an extended \ion{H}{i} disc with a diameter of $226\arcsec$ ($54$~kpc) and a total mass of $6.4\times10^9$~M$_\odot$ (scaled to our adopted distance above).

\begin{table}
  \centering
  \caption{Properties of our ALMA \cotwo{} and $230$-GHz continuum observations. NGC~3393 includes the properties of     \protect\citeauthor{Finlez+2018MNRAS479.3892}'s (\protect\citeyear{Finlez+2018MNRAS479.3892}) observations for reference.}
  \label{tab:observationProperties}
  \setlength\tabcolsep{2pt}
  \begin{tabular}{llc}
    \hline \hline
    Galaxy & Property & Value\\
    \hline
    NGC~1194: & Baseline range (m) & $19$ -- $1808$ \\
           & Maximum recoverable scale (arcsec) & $2.7$ \\
           & On-source time (min.) & $19.75$ \\ 
           & Pixel scale (arcsec~pix$^{-1}$) & $0.05$\\
           & CO channel width (km~s$^{-1}$) & $10$\\
           & CO synthesised beam (arcsec) & $0.30\times0.23$\\
           & CO synthesised beam (pc) & $84\times65$\\
           & CO RMS noise (mJy~beam$^{-1}$~chan$^{-1}$) & $0.41$\\
           & CO integrated flux$^1$ (Jy~km~s$^{-1}$) & $6.1\pm0.1$ \\
           & Integrated molecular gas mass$^2$ (M$_\odot$) & $(5.3\pm0.1)\times10^7$ \\
           & Cont.\ rest-frame frequency (GHz) & $232.1$ \\
           & Cont.\ synthesised beam (arcsec) & $0.31\times0.22$\\
           & Cont.\ synthesised beam (pc) & $87\times62$\\
           & Cont.\ RMS noise (mJy~beam$^{-1}$)& $0.020$\\
    \hline
    NGC~3393:$^3$ & Baseline range (m) & $15$ -- $1100$\\ 
           & Maximum recoverable scale (arcsec) & $5.1$ \\
           & On-source time of (min.) & $49.07$ \\ 
           & Pixel scale (arcsec~pix$^{-1}$) & $0.1$\\
           & CO channel width (km~s$^{-1}$) & $10$\\
           & CO synthesised beam (arcsec) & $0.56\times0.45$\\
           & CO synthesised beam (pc) & $136\times107$\\
           & CO RMS noise (mJy~beam$^{-1}$~chan$^{-1}$) & $0.37$\\
           & CO integrated flux$^1$ (Jy~km~s$^{-1}$) & $81.8\pm0.4$ \\
           & Integrated molecular gas mass$^2$ (M$_\odot$) & $(5.14\pm0.02)\times10^8$ \\
           & Cont.\ rest-frame frequency (GHz) & $238.7$ \\
           & Cont.\ synthesised beam (arcsec) & $0.58\times0.47$\\
           & Cont.\ synthesised beam (pc) & $138\times112$\\
           & Cont.\ RMS noise (mJy~beam$^{-1}$) & $0.017$\\
    \noalign{\vskip 1mm}  
    \hdashline
    \noalign{\vskip 1mm}
    NGC~3393:$^4$ & Baseline range (m) & $15$ -- $629$\\
     & Maximum recoverable scale (arcsec) & $5.3$ \\
           & On-source time (min.) & $28.33$\\
           & CO channel width (km~s$^{-1}$) & $10$\\
           & CO synthesised beam (arcsec) & $0.73\times0.62$\\
           & CO synthesised beam (pc) & $174\times148$\\
           & CO RMS noise (mJy~beam$^{-1}$~chan$^{-1}$) & $0.45$\\
           & Cont.\ rest-frame frequency (GHz) & $239.6$ \\
           & Cont.\ synthesised beam (arcsec) & $0.71\times0.61$\\
           & Cont.\ synthesised beam (pc) & $169\times146$\\
           & Cont.\ RMS noise (mJy~beam$^{-1}$) & $0.023$\\
    \hline
    NGC~5765B: & Baseline range (m) & $15$ -- $1124$\\
           & Maximum recoverable scale (arcsec) & $3.5$ \\
           & On-source time (min.) & $22.25$\\ 
           & Pixel scale (arcsec~pix$^{-1}$) & $0.06$\\
           & CO channel width (km~s$^{-1}$) & $10$\\
           & CO synthesised beam (arcsec) & $0.47\times0.28$\\
           & CO synthesised beam (pc) & $257\times153$\\
           & CO RMS noise (mJy~beam$^{-1}$~chan$^{-1}$) & $0.48$\\
           & CO integrated flux$^1$ (Jy~km~s$^{-1}$) & $216.2 \pm 0.4$ \\
           & Integrated molecular gas mass$^2$ (M$_\odot$) & $(6.94\pm0.01)\times10^9$ \\
           & Cont.\ rest-frame frequency (GHz) & $232.5$ \\
           & Cont.\ synthesised beam (arcsec) & $0.47\times0.28$\\
           & Cont.\ synthesised beam (pc) & $257\times153$\\
           & Cont.\ RMS noise (mJy~beam$^{-1}$) & $0.028$\\
    \hline \hline
  \end{tabular}
  \parbox{\columnwidth}{\scriptsize
    $^1$ The CO integrated fluxes and associated integrated molecular gas masses are measured within the ALMA FoV only, and thus likely do not cover the full molecular gas extent of NGC~1194 and NGC~3393.\\
    $^2$ A \cotwo{}/\coone{} line ratio of unity (in brightness temperature units) and a CO-to-molecule conversion factor of $4.3$~M$_{\odot}$~pc$^{-2}$~(K~km~s$^{-1}$)$^{-1}$ are assumed to infer the molecular gas masses, that include the contribution of heavy elements.\\
    $^3$ All the quantities listed in this segment are measured after combining the observations from programmes 2016.1.01553.S and 2015.1.00086.S.\\
    $^4$ All the quantities listed in this segment are taken from \citet{Finlez+2018MNRAS479.3892}.}
\end{table}

\subsection{NGC~5765B}

NGC~5765B is an Sab galaxy \citep{Pjanka+2017ApJ844.165} with a Seyfert~2 nucleus \citep{Toba+2014ApJ788.45}. It has a close companion, NGC~5765A, at an angular separation of $22\farcs5$ ($\approx12$~kpc at our adopted distance below), beyond the FoV of our ALMA and JCMT observations. \textit{HST} imaging reveals a bar, two rings (at radii of $1\farcs5$ and $3\farcs5$) and spiral features between these rings. Beyond the outer ring the galaxy is perturbed by interaction with its companion \citep{Pjanka+2017ApJ844.165}.

Central megamasers were detected with the Green Bank Telescope as part of the MCP and were monitored for over two years, yielding an angular diameter distance $D=112\pm5$~Mpc (adopted here; \citealt{2020ApJ...891L...1P})
and a SMBH mass $M_\mathrm{BH}=(4.0\pm0.4)\times10^7$~M$_\odot$ (at this adopted distance; \citealt{Gao+2016ApJ817.128}). 
The maser disc ($\approx1.2$~pc in radius) warps over an inclination range of $i=94\fdg5$ at the centre to $i=73\fdg3$ at the edge and over a kinetic position angle range of $PA_\mathrm{kin}=146\fdg7$ at the centre to $PA_\mathrm{kin}=139\fdg8$ at the edge, as determined through dynamical modelling \citep{Gao+2016ApJ817.128}. The overall galaxy inclination is $i=26\degr$ with $PA_\mathrm{mor}=10\degr$, as determined from a 2MASS $K_\mathrm{s}$-band image \citep{2006AJ....131.1163S}.

\citet{2012MNRAS.421.1043S} studied the ionised gas of NGC5765B using an optical spectrum from SDSS. They reported strong nebular \ion{He}{ii} emission lines dominated by the AGN (i.e.\ without Wolf–Rayet features), along with other emission lines. The Arecibo Legacy Fast Arecibo L-band Feed Array survey \citep{2018ApJ...861...49H} derived a neutral hydrogen content of $(6.3\pm0.7)\times10^9$~M$_\odot$ (at our adopted distance above). \citet{2022MNRAS.512.1522D} presented an ALMA \cotwo{} map and reported a total mass of $(1.2\pm0.1)\times10^{10}$~M$_\odot$ (at our adopted distance above).

\section{Observations}
\label{sec_observations}

\subsection{ALMA observations}
\label{sec_alma}

Observations of the \cotwo{} line of our three target galaxies were carried out with ALMA as part of programme 2016.1.01553.S (PI: Bureau) on 24 October 2016 and 2, 3 and 15 May 2017. For NGC~3393, we combine another track from ALMA programme 2015.1.00086.S (PI: Nagar) obtained on 3 May 2016.

\begin{figure*}
  \includegraphics[width=\textwidth]{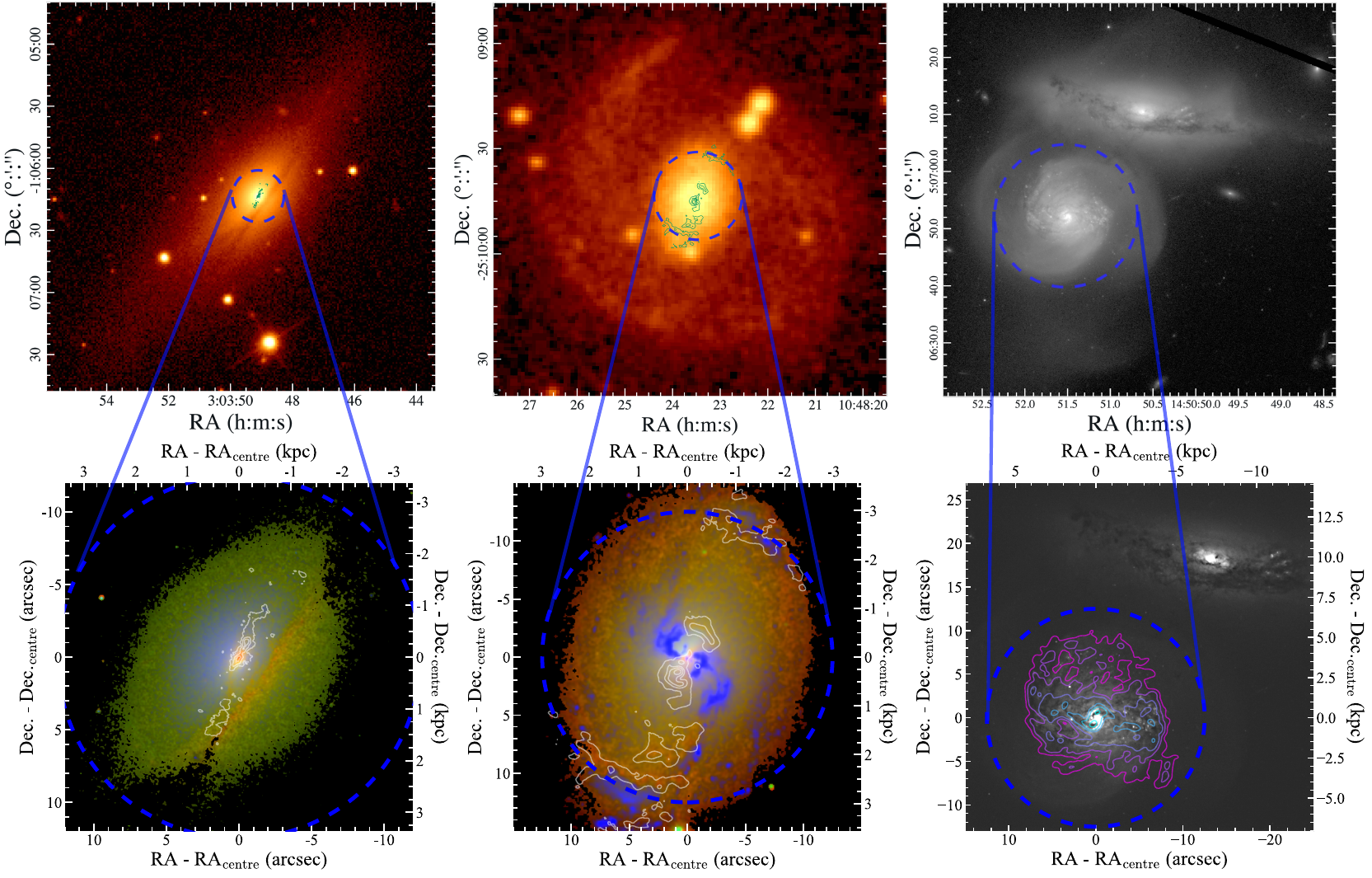}
  \caption{Optical images of our three target galaxies. In each panel the ALMA \cotwo{} emission is overlaid as contours (except the upper-right panel) and the ALMA primary beam is shown as a blue dashed circle. The bottom row panels show zoom-in images.
  {\bf Left column:} NGC~1194. Upper panel: SDSS $r$-band image of the whole galaxy. Lower panel: \textit{HST} Wide-field Camera~3 (WFC3) F438W (blue), F814W (green) and F160W (red) composite of the central region.
  {\bf Middle column:} NGC~3393. Upper panel: Digitized Sky Survey-2 red-band image of the whole galaxy. Lower panel: \textit{HST}/WFC3 F336W (blue), F814W (green) and F160W (red) composite of the central region. 
  {\bf Right column:} NGC~5765B (in the bottom-left corner of each panel) and its companion NGC~5765A. Upper panel: \textit{HST}/WFC3 F814W image. Lower panel: unsharp-masked \textit{HST}/WFC3 F814W image.
  }
  \label{fig:optOver}
\end{figure*}

For all our observations, the ALMA correlator was configured with one spectral window centred on the redshifted frequency of the \cotwo{} line (rest frequency $\nu_\mathrm{rest}=230.538$~GHz.), with a bandwidth of $1.875$~GHz ($2438$~km~s$^{-1}$ at $z=0$) and $488$-kHz channels ($0.63$~km~s$^{-1}$ at $z=0$). The remaining three spectral windows were used to observe the continuum, if any, each with a bandwidth of $2$~GHz and $15.625$-MHz channels. The additional track for NGC~3393 also had one spectral window centred on \cotwo{}, with an additional spectral line window centred on CS(5-4) and two continuum spectral windows. Details of this track can be found in \citet{Finlez+2018MNRAS479.3892}. The FoV of the ALMA 12-m array, i.e.\ the FWHM of the ALMA 12-m antennae primary beam, is $\approx25\arcsec$ ($\approx7.0$, $6.0$ and $15$~kpc for NGC~1194, NGC~3393 and NGC~5765B, respectively) at the observed frequencies of \cotwo{}. This FoV only extends to $1.1$~$R_\mathrm{e}$ in NGC~1194 ($H$-band combined with $K$-band; \citealt{Lasker+2016ApJ825.3}), $1.3$~$R_\mathrm{e}$ in NGC~3393 ($H$-band combined with $I$-band; \citealt{Lasker+2016ApJ825.3}) and $1.8$~$R_\mathrm{e}$ in NGC~5765B ($K_\mathrm{s}$-band; \citealt{2022MNRAS.512.1522D}), where $R_\mathrm{e}$ is the effective (i.e.\ half-light) radius. The ALMA FoV therefore covers the whole galaxy disc only in NGC~5765B, with more limited coverage in NGC~1194 and NGC~3393 (see \autoref{fig:optOver})

The baseline ranges of the observations of NGC~1194, NGC~3393 and NGC~5765B were $19$ -- $1808$, $15$ -- $1100$ and $15$ -- $1124$~m, respectively. The corresponding maximum recoverable scales were $2\farcs7$ ($0.86$~kpc), $5\farcs1$ ($1.2$~kpc) and $3\farcs5$ ($2.1$~kpc), respectively. The data were calibrated using the standard ALMA pipeline, either through the European ALMA Regional Centre Calibrated Measurement Set (CalMS) service or by locally running \texttt{Common Astronomy   Software Applications}\footnote{Available from \url{https://casa.nrao.edu/}.} ({\tt CASA}; \citealt{2022PASP..134k4501C}). CASA version~4.7 was used to calibrate the three tracks of 2016.1.01553.S and version~4.6 for the track of 2015.1.00086.S. For NGC~3393, the ALMA observations of \cite{Finlez+2018MNRAS479.3892} were combined with our higher-angular resolution observations to improve the $uv$-plane coverage and sensitivity (see \autoref{tab:observationProperties}). For both the line datacubes and the continuum images, the combined data improve the angular resolution (in one dimension) by $\approx25\%$ compared to \cite{Finlez+2018MNRAS479.3892}. The following imaging steps all used CASA version~6.4.4.

First, for each galaxy, continuum emission was subtracted from the visibility data using linear fits and the CASA task {\tt uvcontsub}. To produce datacubes with high sensitivity, we binned the channels to $10$~km~s$^{-1}$ and used  Briggs weighting with a robust parameter of $2.0$ (close to natural weighting). The datacubes were then cleaned to a depth of $1.5$ times the root-mean-square (RMS) noise of the line-free channels using the task {\tt tclean} with the {\tt MultiScale} algorithm \citep{2008ISTSP...2..793C} in the interactive mode and using a manually-defined three-dimensional mask. We note that varying these parameters does not significantly change any result of this paper. A primary beam correction was then applied to the line datacubes.

Moment maps were generated using a smooth-masking technique \citep{Dame2011arXiv1101.1499}. A smoothed datacube (without primary-beam correction) was first generated by spatially convolving every channel with a two-dimensional (2D) square uniform kernel of side length equal to the synthesised beam width. As the channels are already binned to a width of $10$~km~s$^{-1}$, we did not smooth further spectrally. We then constructed a mask by first selecting all pixels of the smoothed datacube above a given flux density threshold ($1.5$~RMS) and then excluding pixels outside of the mask manually defined during cleaning. Finally, the mask was adjusted channel by channel by (i) filling `holes' of unselected pixels with areas smaller than two synthesised beams and (ii) removing `islands' of selected pixels with areas smaller than one synthesised beam (both achieved using the Python package {\tt scikit-image}\footnote{Available from \url{https://scikit-image.org}.}; \citealt{2014PeerJ...2..453V}). This mask was then used to select the pixels of the original unsmoothed primary beam-corrected datacube that are used for the moment analysis.

\begin{landscape}
  \begin{figure}
    \begin{center}
      \includegraphics[width=1.25\textwidth]{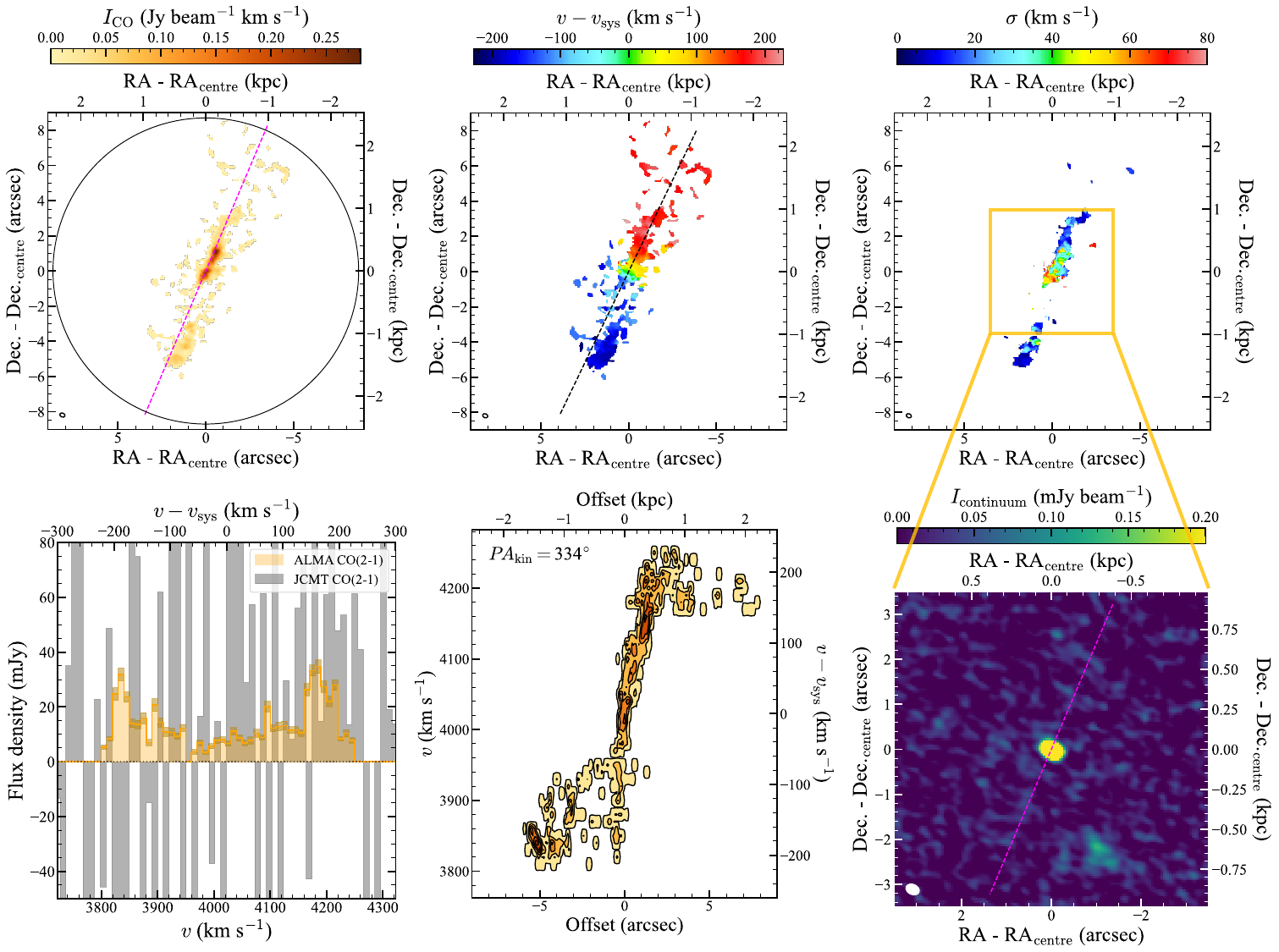}
      \caption{Data products of NGC~1194. \textbf{Top-left:}
        \cotwo{} zeroth-moment (integrated flux) map. The overlaid black solid circle shows the JCMT beam (assuming perfect pointing accuracy) at the same frequency. The magenta dashed line indicates the position angle of the maser disc (extrapolated in spatial extent). \textbf{Top-centre:} \cotwo{} first-moment (intensity-weighted mean velocity) map. The overlaid black dashed line shows the kinematic major axis. \textbf{Top-right:} \cotwo{} second-moment (intensity-weighted velocity dispersion) map. \textbf{Bottom-left:} \cotwo{} integrated spectrum synthesised from the ALMA datacube (orange histogram, with uncertainties indicated as darker shades) and JCMT spectrum (grey histogram). \textbf{Bottom-centre:} \cotwo{} PVD along the kinematic major axis, whose position angle is listed in the top-left corner. \textbf{Bottom-right:} $230$-GHz continuum map. The magenta dashed line again indicates the position angle of the maser disc (extrapolated in spatial extent). The spatial extent shown is smaller than that of the moments maps, as illustrated by the overlaid orange square. The synthesised beam is shown in the bottom-left corner of all the maps.}
      \label{fig:NGC1194_maps}
    \end{center}
  \end{figure}
\end{landscape}

\begin{landscape}
  \begin{figure}
    \begin{center}
      \includegraphics[width=1.25\textwidth]{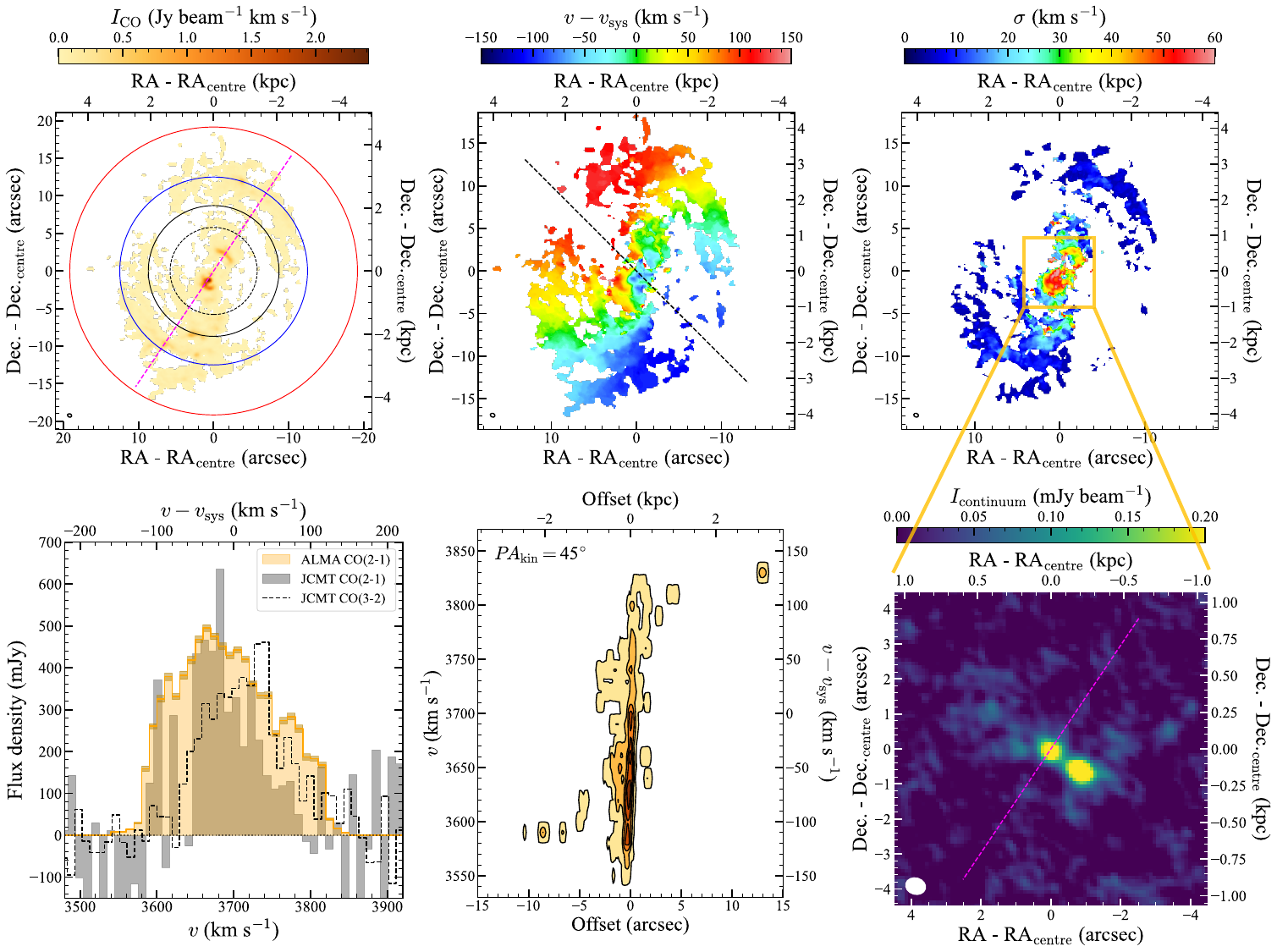}
      \caption{As \autoref{fig:NGC1194_maps} but for NGC~3393. The JCMT beam (assuming perfect pointing accuracy) at the \cothree{} frequency is also shown in the top-left panel as a black dashed circle, and the corresponding spectrum is shown in the bottom-left panel as a black dashed histogram. The ALMA \cotwo{} primary beam full width at $50\%$ (i.e.\ the usual primary beam definition) and at $20\%$ (the maximal extent of the datacube) of the maximum are also shown in the top-left panel as blue and red solid circles, respectively. Due to the scarcity of gas along the kinematic major axis, the kinematic position angle of $45\degr$ was estimated by eye rather than by a fit. The \cotwo{} emission reaches the edge of the ALMA FoV.}
      \label{fig:NGC3393_maps}
    \end{center}
  \end{figure}
\end{landscape}

\begin{landscape}
  \begin{figure}
    \begin{center}
      \includegraphics[width=1.25\textwidth]{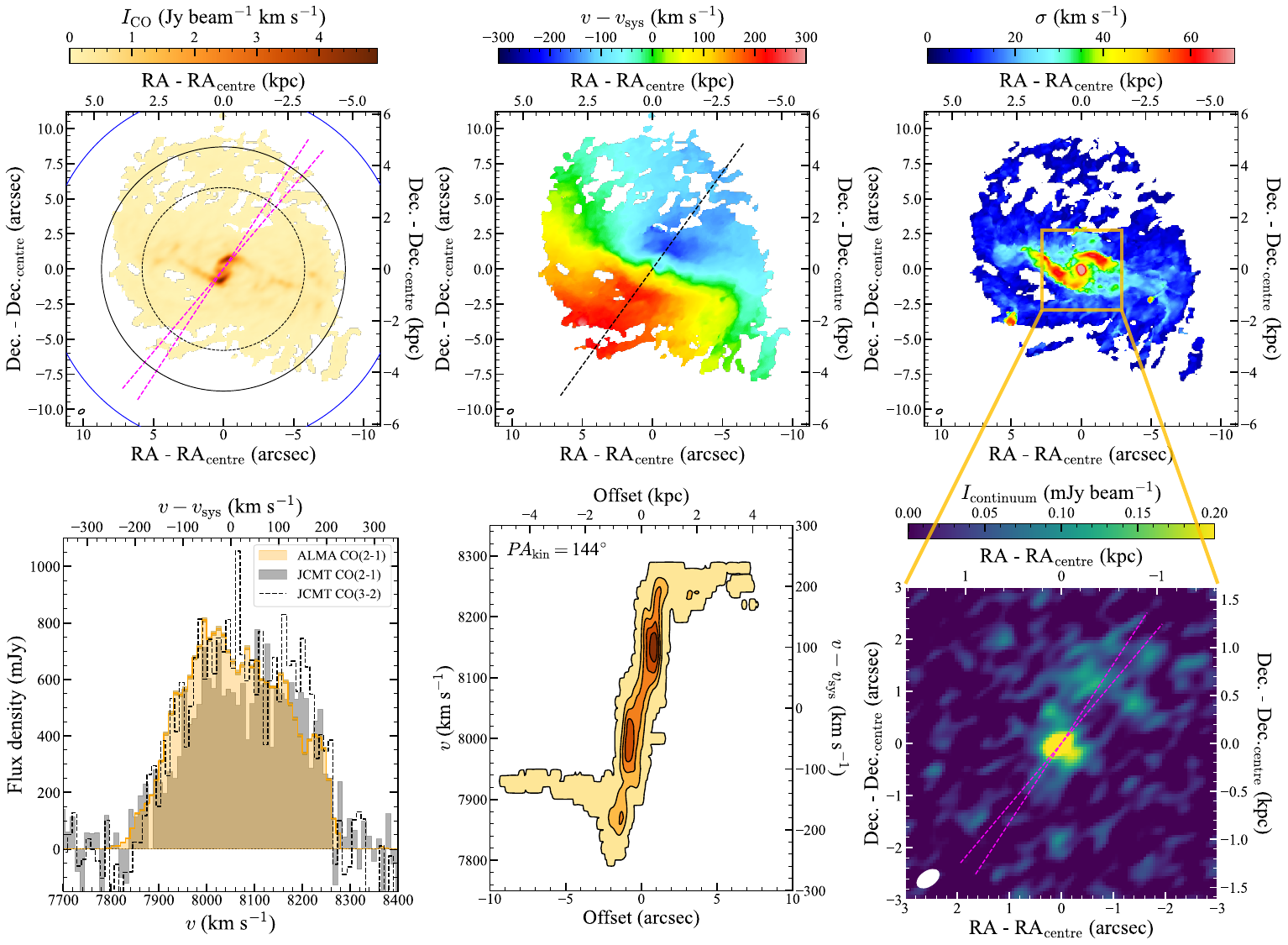}
      \caption{As \autoref{fig:NGC3393_maps} but for NGC~5765B (although the red solid circle is not visible and the range of position angles of the maser disc due to its warp is indicated by two magenta dashed lines in the relevant panels).}
      \label{fig:NGC5765B_maps}
    \end{center}
  \end{figure}
\end{landscape}

The zeroth-moment (integrated flux), first-moment (intensity-weighted mean velocity) and second-moment\footnote{The second-moment requires at least two selected pixels along any line of sight, explaining the difference of spatial coverage between the second-moment map and other moment maps.} (intensity-weighted velocity dispersion) maps of NGC~1194, NGC~3393 and NGC~5765B are shown in the top rows of \autoref{fig:NGC1194_maps}, \autoref{fig:NGC3393_maps} and \autoref{fig:NGC5765B_maps}, respectively. The same mask was used to create the integrated spectrum (and to calculate the integrated flux) and the kinematic major-axis position-velocity diagram (PVD) of each galaxy, shown in the bottom-left and bottom-centre panel of each figure, respectively. As each channel has a different 2D masked region and spaxels near each other are strongly correlated, we applied the following procedure to estimate the uncertainty of the spatially integrated flux of each channel. For each line channel, we applied its 2D mask to every line-free channel of the datacube, calculated the sum of the flux densities of each of those line-free channels and adopted the RMS (around the mean) of these sums as the uncertainty of the integrated flux of the original line channel. Standard error propagation rules\footnote{Given the raw spectral resolution and the binned channel width, we safely assume no covariance between adjacent channels.} were then used to calculate the uncertainty of the flux integrated over all channels (and in turn of the total molecular gas mass). Each PVD was generated by adopting a $2\arcsec$-wide mock slit along the kinematic major axis, determined from a fit to the first-moment map using the {\tt pafit} software\footnote{Available from \url{https://pypi.org/project/pafit/}.} \citep{Krajnovic2006}, except for NGC~3393 for which a rough kinematic major axis position angle was estimated by visual inspection due to the scarcity of gas along it. We used the astrometry from the existing VLBI observations to define the centre of each galaxy (see \autoref{tab:targets}).

For each galaxy, a continuum image was also produced from all spectral windows using only line-free channels. With the {\tt tclean} task in the multi-frequency synthesis mode, we again adopted Briggs weighting with a robust parameter of $2.0$ and cleaned to a depth of $1.5$ times the RMS noise of the dirty image (measured in emission-free regions) in the interactive mode and using a manually-defined 2D mask. The primary beam correction was then applied to the continuum images.

The final angular resolution, sensitivity and other characteristics of each data product are listed in \autoref{tab:observationProperties}.

\subsection{JCMT observations}

Single-dish \cotwo{} observations of NGC~1194, NGC~3393 and NGC5765B were obtained with the JCMT on 18 -- 23 October 2016, as part of Programme M16BP068 (PI: Gao), aiming to estimate the prevalence and abundance of molecular gas in the $15$ disc galaxies known to harbour megamasers at the time and another $15$ galaxies used as a control sample (Gao et al.\ in prep.). Follow-up \cothree{} observations of NGC~3393 and NGC~5765B were carried out on 6 July 2017, 5 December 2017, 22 January 2018, 29 January 2018 and 24 December 2018, as part of Programme M17BP056 (PI: Gao).

The JCMT has a $15$-m diameter antenna and therefore beams (i.e.\ FoVs) of $17\farcs4$ and $11\farcs6$ at the rest frequency of \cotwo{} and of \cothree{}, respectively, both of which are smaller than the ALMA primary beam in this work. \autoref{tab:JCMTproperties} lists the basic properties of the JCMT observations, while the \cotwo{} and \cothree{} JCMT beams are shown in Figures~\ref{fig:NGC1194_maps} -- \ref{fig:NGC5765B_maps} as black solid and black dashed circles, respectively, overlaid on the zeroth-moment maps.

\begin{table}
  \centering
  \caption{Properties of our JCMT \cotwo{} and \cothree{} observations.}
  \label{tab:JCMTproperties}
  \begin{tabular}{llc}
    \hline \hline
    Galaxy & Property & Value\\
    \hline
    NGC~1194 & Velocity range (km~s$^{-1}$) & 3800 -- 4300 \\
           & \cotwo{} RMS noise (mJy) & $83$ \\
           & \cotwo{} int.\ flux (Jy~km~s$^{-1}$) & $<17.8$ \\
           & $M_\mathrm{mol}$~(M$_\odot$) & $<1.55\times10^8$ \\
    \hline
    NGC~3393 & Velocity range (km~s$^{-1}$) & 3580 -- 3900 \\
         & \cotwo{} RMS noise (mJy) & 131 \\
         & \cotwo{} int.\ flux (Jy~km~s$^{-1}$) & $43\pm10$ \\
         & $M_\mathrm{mol}$~(M$_\odot$) & ($2.7\pm0.6)\times10^8$ \\
         & \cothree{} RMS noise (mJy) & 71 \\
         & \cothree{} int.\ flux (Jy~km~s$^{-1}$) & $52\pm5$ \\
         & JCMT \cotwo{}/\cothree{} (K/K) & $1.8\pm0.5$ \\
    \hline
    NGC~5765B & Velocity range (km~s$^{-1}$) & 7750 -- 8400 \\
         & \cotwo{} RMS noise (mJy) & 115 \\
         & \cotwo{} int.\ flux (Jy~km~s$^{-1}$) & $186\pm16$ \\
         & $M_\mathrm{mol}$~(M$_\odot$) & $(6.0\pm0.5)\times10^9$ \\
         & \cothree{} RMS noise (mJy) & 67 \\
         & \cothree{} int.\ flux (Jy~km~s$^{-1}$) & $207\pm14$ \\
         & JCMT \cotwo{}/\cothree{} (K/K) & $2.02\pm0.22$ \\
    \hline \hline
  \end{tabular}
    \parbox{\columnwidth}{
  Note: The integrated CO fluxes and associated integrated molecular gas masses and CO line ratios are only measured within the JCMT FoVs, that do not necessarily cover the full extent of the galaxies' molecular gas. The RMS noises were measured using a channel width of $10$~km~s$^{-1}$.}
\end{table}

For the \cotwo{} observations, the RxA receiver was used with dual sidebands, with the Auto Correlation Spectral Imaging System (ACSIS) correlator configured to have a bandwidth of $1$~GHz and channels of $1$~MHz ($\approx1.3$~km~s$^{-1}$ at $z=0$). The observations were conducted in standard position-switching mode with a total on-source integration time of $\approx30$~min per target. For the \cothree{} observations, the $16$-pixel Heterodyne Array Receiver Program (HARP) was used, with the ACSIS correlator configured to have a bandwidth of $1$~GHz and channels of $0.488$~MHz ($\approx0.4$~km~s$^{-1}$ at $z=0$). The observations were conducted in beam-switching mode to achieve a better spectral baseline, with a total on-source integration time of $\approx1$~hr per target.

For both \cotwo{} and \cothree{} data reduction, the standard ORAC data reduction pipeline (ORAC-DR; \citealt{2015A&C.....9...40J}) was used within the \texttt{Starlink} software package\footnote{Available from \url{https://starlink.eao.hawaii.edu/starlink}\,.} \citep{2014ASPC..485..391C}. The spectra were first rebinned to channels of $10$~km~s$^{-1}$ to match our ALMA data. A linear baseline fit to line-free channels was then subtracted from each spectrum, to remove any continuum emission and non-flat baseline. The spectra were then converted to a flux density scale $S_\mathrm{v}=15.6\,T_\mathrm{A}^\star/\eta_\mathrm{A}$, where $T_\mathrm{A}^\star$ is the antenna temperature and $\eta_\mathrm{A}=0.52$ is the JCMT antenna efficiency at both frequencies. The resulting spectra of NGC~1194, NGC~3393 and NGC~5765B are shown in the bottom-left panels of \autoref{fig:NGC1194_maps}, \autoref{fig:NGC3393_maps} and \autoref{fig:NGC5765B_maps}, respectively.

The integrated flux of each spectrum was obtained by integrating over the velocity range with clear emission (or a range estimated from the ALMA data in the case of  NGC~1194, which is a non-detection), as listed in \autoref{tab:JCMTproperties}, while the integrated flux uncertainty is estimated using
\begin{equation}
   \sigma=\Delta v\,\sigma_\mathrm{chan}\,\sqrt{N_\mathrm{line}\left(1+\frac{N_\mathrm{line}}{N_\mathrm{noise}}\right)}\,\,\,,
\end{equation}
where $\Delta v$ is the channel width ($10$~km~s$^{-1}$), $\sigma_\mathrm{chan}$ the noise per channel listed in \autoref{tab:JCMTproperties}, $N_\mathrm{line}$ the number of channels used for the integration and $N_\mathrm{noise}$ the number of channels used to estimate the noise \citep[see][]{Young+2011MNRAS414.940}. As for the ALMA data, the molecular gas mass of each galaxy was then calculated from the \cotwo{} line by assuming a \cotwo{}/\coone{} line ratio $R_{21}$ of unity (in brightness temperature units) and a CO-to-molecule conversion factor $\alpha_\mathrm{CO}=4.3$~M$_\odot$~pc$^{-2}$~(K~km~s$^{-1}$)$^{-1}$ \citep{2013ARA&A..51..207B}, including the contribution of heavy elements, yielding
\begin{equation}
    \frac{M_\mathrm{mol}}{\mathrm{M}_\odot}=\frac{2.63\times10^3}{1+z_{\rm helio}}\,\left(\frac{F_\mathrm{\cotwo{}}}{\mathrm{Jy~km~s^{-1}}}\right)\,\left(\frac{D}{\mathrm{Mpc}}\right)^{2}\,\,\,,
\end{equation}
where $F_\mathrm{\cotwo{}}$ is the intregrated \cotwo{} flux density.

\subsection{NGC~1194}
\label{sec:1194_obs}

\subsubsection{Molecular gas}
\label{sec:1194_co}

The moment maps, PVD and integrated spectrum of NGC~1194 shown in \autoref{fig:NGC1194_maps} suggest an edge-on, disturbed and lopsided central molecular gas disc well aligned with the maser disc. There are two large molecular gas concentrations, one extending north-west from the centre, the other farther out and disconnected to the south-east. Comparison to the SDSS $r$-band image in the upper-left panel of \autoref{fig:optOver} reveals the concentrations to be roughly aligned with the galaxy morphological major axis, while the composite \textit{HST} image (F438W, F814W and F160W filters) in the lower-left panel of \autoref{fig:optOver} shows the molecular gas to be associated primarily with a weak irregular dust lane crossing the galaxy centre (with $PA_\mathrm{mor}\approx160\degr$), while a much more prominent dust lane (offset to the south-west with $PA_\mathrm{mor}\approx145\degr$) that seems to define part of a dust ring is apparently devoid of CO emission. This morphology and the first-moment map suggest the molecular gas to be part of an irregular structure embedded within a roughly edge-on rotating disc, itself aligned with the large-scale galaxy disc. Unsurprisingly, the second-moment map reveals irregular velocity dispersions. The velocity dispersions are highest in the centre, but this may be due to beam smearing. They decrease to $10$ -- $20$~km~s$^{-1}$ at the extremities of the distribution, typically associated with dynamically cold gas, and are comparable to the velocity dispersions of discs reported in other works \citep[e.g.][]{Davis+2018MNRAS473.3818, Smith+2019MNRAS485.4359, 2022MNRAS.516.4066L,  2022MNRAS.514.5035L}. Overall, our observations suggest a disturbed molecular gas distribution in a non-equilibrium state, possibly caused by an earlier (minor) merger event, an hypothesis supported by the twisted morphological position angle at large radii (see Fig.~11 of \citealt{Lasker+2016ApJ825.3}) as well as the existence of a detached \ion{H}{i} cloud to the northwest of the main galaxy disc \citep{Sun+2013ApJ778.47}. Despite this, the maximum rotation velocity matches well that measured in \ion{H}{i} \citep{Sun+2013ApJ778.47}.

As shown in the bottom-left panel of \autoref{fig:NGC1194_maps}, NGC~1194 is not detected in our JCMT \cotwo{} observations, that have a sensitivity of only $83$~mJy per channel (while our synthesised ALMA integrated spectrum has a sensitivity of $1.3$~mJy per channel on average, both based on the same channel width of $10$~km~s$^{-1}$). The ALMA integrated flux of $6.1\pm0.1$~Jy~km~s$^{-1}$ yields a total molecular gas mass of ($5.3\pm0.1)\times10^7$~M$_\odot$. Here and for the other two galaxies, the uncertainties quoted on the integrated flux and associated integrated mass are exclusively due to the noise in the integrated spectrum (see \autoref{sec_alma}). They do not include potential systematic errors such as the ALMA flux calibration uncertainty (typically $\approx10\%$), CO-to-molecule conversion factor uncertainty (typically $\approx30\%$; \citealt{2013ARA&A..51..207B}), galaxy distance uncertainty (typically $\approx10\%$), etc. As the ALMA and JCMT FoVs only encompass $\approx1$~$R_\mathrm{e}$ around the galaxy centre, we have no information on the molecular gas beyond that region, and both the integrated flux and the associated integrated mass calculated here are likely lower limits of those quantities for the entire galaxy.

\subsubsection{230-GHz continuum emission}

The continuum map shown in the bottom-right panel of \autoref{fig:NGC1194_maps} reveals a single compact $230$-GHz continuum source at the centre of NGC~1194, most likely associated with the AGN. Fitting this source with a 2D Gaussian using the {\tt CASA} task {\tt imfit} reveals it to be slightly spatially resolved (i.e.\ slightly larger than the synthesised beam). Its position and flux density derived from the Gaussian fit are listed in \autoref{tab:pointsources}. This position is marginally consistent with that derived from maser astrometry (see \autoref{tab:targets}). Another tentative more diffuse source is located $\approx2\farcs4$ south-west of the centre, with an integrated flux of $\approx0.55$~mJy.

\begin{table}
  \centering
  \caption{Properties of compact $230$-GHz sources within the fields of view of our observations.}
  \label{tab:pointsources}
  \begin{tabular}{llc} 
    \hline \hline
    Galaxy & Property & Value \\
    \hline
    NGC~1194 & R.A.\ (J2000) & $03^\mathrm{h}03^\mathrm{m}49\fs109$ \\
           & Dec.\ (J2000) & $-1\degr06\arcmin13\farcs48$ \\
           & Flux (mJy) & $1.73\pm0.04$ \\
    \hline
    NGC~3393 (nucleus) & R.A.\ (J2000) & $10^\mathrm{h}48^\mathrm{m}23\fs47$ \\
           & Dec.\ (J2000) & $-25\degr09\arcmin43\farcs5$ \\
           & Flux (mJy) & $0.40\pm0.04$ \\
           & Spectral index & $-0.18\pm0.03$ \\
    NGC~3393 (SW) & R.A.\ (J2000) & $10^\mathrm{h}48^\mathrm{m}23\fs40$ \\
           & Dec.\ (J2000) & $-25\degr09\arcmin44\farcs1$ \\
           & Flux (mJy) & $0.60\pm0.04$ \\
           & Spectral index & $-0.8\pm0.3$ \\
    \hline
    NGC~5765B (nucleus) & R.A.\ (J2000) & $14^\mathrm{h}50^\mathrm{m}51\fs52$ \\
           & Dec.\ (J2000) & $+5\degr06\arcmin52\farcs2$ \\
           & Flux (mJy) & $0.71\pm0.08$ \\
    NGC~5765B (SW) & RA (J2000) & $14^\mathrm{h}50^\mathrm{m}51\fs50$ \\
           & Dec.\ (J2000) & $+5\degr06\arcmin51\farcs9$ \\
           & Flux (mJy) & $0.28\pm0.08$ \\
    \hline
    \hline
  \end{tabular}
  \parbox{0.45\textwidth}{Notes: Source positions and integrated fluxes were measured using Gaussian fits. The spectral indices of NGC~3393 were measured by cross-identifying the sources with Very Large Array observations by \protect\cite{Koss+2015ApJ807.149} and fitting power laws of the form $S\propto\nu^\alpha$ to the spectral energy distributions, where $S$ is the integrated flux density, $\nu$ the frequency and $\alpha$ the spectral index.}
\end{table}

\subsection{NGC~3393}
\label{sec:3393_obs}

\subsubsection{Molecular gas}

The moments maps, PVD and integrated spectra of NGC~3393 shown in \autoref{fig:NGC3393_maps} suggest a fairly regular but patchy molecular gas distribution, with little gas near the kinematic major axis and only faint diffuse gas in the very centre. 
There are two brighter structures in the central region, one south-east of the centre ($\approx1\farcs5$ or $360$~pc from the centre), the other to the north-west ($\approx3\farcs3$ or $790$~pc from the centre), whose morphologies are reminiscent of (part of) a nuclear ring and/or spiral. These are also associated with increased velocity dispersions ($40$ -- $60$~km~s$^{-1}$), that are otherwise ordinary ($10$ -- $20$~km~s$^{-1}$). The velocity field is fairly regular on large scales, with $PA_\mathrm{kin}\approx45\degr$, although a kinematic twist is present in the outer parts (most easily seen as a clear kink in the zero-velocity curve) and there are many small-scale disturbances. \citet{Finlez+2018MNRAS479.3892} discussed the observed \cotwo{} kinematics in great detail, along with the stellar and ionised-gas kinematics. To explain both the large-scale kinematics and that near the two brighter sources, they proposed a perturbation model driven by both a large-scale bar and a nuclear bar. The molecular gas detected at the largest radii ($\approx10\arcsec$) forms an annulus or ring-like structure, that may be associated with spiral arms observed in the UV (see the lower-middle panel of \autoref{fig:optOver}), and may thus trace recent star formation. Finally, the lack of molecular gas along the kinematic major axis may be due to photo-ionisation by the AGN/jets detected in continuum emission (see below), that are perpendicular to the accretion disc traced by maser emission (magenta lines in \autoref{fig:NGC3393_maps}).

As shown in the bottom-left panel of \autoref{fig:NGC3393_maps}, our JCMT \cotwo{} spectrum is in good agreement with our synthesised ALMA integrated spectrum at velocities above $3650$~km~s$^{-1}$, but the two differ at smaller velocities, with much lower JCMT fluxes. The ALMA first-moment map indicates that this could be accounted for if the JCMT had a pointing offset. 
Independent of this, despite the fact that our
ALMA data do not have baselines shorter than $15$~m, ALMA generally recovers more flux than the JCMT, likely because of the smaller FoV of the latter and the extended molecular gas distribution. The integrated flux of our ALMA \cotwo{} cube is $81.8\pm0.4$~Jy~km~s$^{-1}$, yielding a total molecular gas mass of $(5.14\pm0.02)\times10^8$~M$_\odot$. As the ALMA FoV only encompasses $\approx1.3$~$R_\mathrm{e}$ and the molecular gas clearly extends to the FoV's edge, both the integrated flux and the associated integrated mass should again be considered lower limits of those quantities for the entire galaxy.

Our \cothree{} JCMT spectrum is also in good agreement with our synthesised ALMA integrated \cotwo{} spectrum at velocities above $3720$~km~s$^{-1}$, but it again shows a significant flux deficit at lower velocities. This could again be explained by a pointing offset. The integrated JCMT \cothree{} flux of $52\pm5$~Jy~km~s$^{-1}$ is thus both highly unreliable and likely a lower limit.

\subsubsection{230-GHz continuum emission}

The $230$-GHz continuum map shown in the bottom-right panel of \autoref{fig:NGC3393_maps} reveals two compact sources, one at the galaxy centre, consistent with the VLBI maser location, that we will refer to as the nuclear source, the other offset by $\approx1\farcs0$ or $\approx240$~pc south-west of the nucleus. Both sources are marginally spatially resolved, and the south-west source is $\approx50\%$ brighter than the nuclear source. There is also faint fuzzy emission north-east of the nucleus.

Based on X-ray emission, \citet{Fabbiano+2011Nature477.431} reported a pair of SMBH/AGN separated by $150$~pc ($0\farcs6$), much smaller than the separation of $\approx240$~pc between the two $230$-GHz compact sources discussed above. Comparing the positions of the two X-ray sources in Figure~1 of  \citet{Fabbiano+2011Nature477.431}, they appear to both be located in the emission tail of our nuclear source, both far away  from the south-west compact source. Therefore, our two compact sources are unlikely to be \citeauthor{Fabbiano+2011Nature477.431}'s (\citeyear{Fabbiano+2011Nature477.431}) claimed dual SMBH/AGN. We also note that while \citet{Finlez+2018MNRAS479.3892} also disfavoured a SMBH/AGN pair, they misreported the location of the claimed second source in their Figure~4 and Section~4.3 which, according to Figure~1a of \citet{Fabbiano+2011Nature477.431}, should be north-east of the peak of the Very Large Array's (VLA) $8.4$-GHz central source.

The nuclear source overlaps with component A discovered by \citeauthor{Koss+2015ApJ807.149} (\citeyear{Koss+2015ApJ807.149}; also reported by \citealt{Finlez+2018MNRAS479.3892}) using VLA $8.4$- and $4.9$-GHz continuum observations, while the south-west compact source partially overlaps with their component B. A third source north-east of the centre was also reported by both \citet{Koss+2015ApJ807.149} as their component C and by \citet{Finlez+2018MNRAS479.3892}. By carefully checking the spatial extent of component C, we conclude that the fuzzy emission detected here north-east of the nucleus also overlaps with it.

\begin{figure}
  \begin{center}
    \centering
    \includegraphics[width=0.4\textwidth]{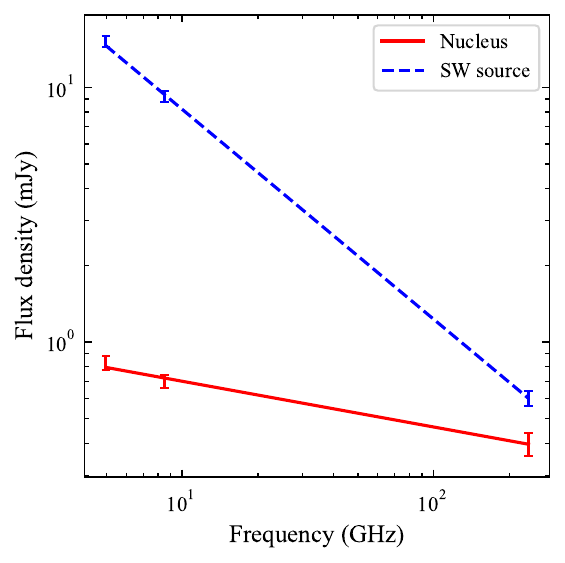}
    \caption{Radio -- millimetre integrated continuum flux densities and power-law fits of the nuclear source (red) and south-west compact source (blue) detected in NGC~3393. The $8.4$- and $4.9$-GHz measurements are from \protect\cite{Koss+2015ApJ807.149}.}
    \label{fig:NGC3393_continuum}
  \end{center}
\end{figure}

These three sources (nuclear source, south-west compact source and north-east fuzzy emission) can be attributed to the central AGN and (intrinsically) symmetric jets on both sides of it. The south-west compact source is associated with the approaching jet and is thus significantly brighter than the north-east fuzzy emission (associated with the receding jet) due to Doppler-boosting \citep{Koss+2015ApJ807.149}. The integrated fluxes of the two compact sources are reported in \autoref{tab:pointsources}, again derived using 2D Gaussian fits carried out with the {\tt CASA} task {\tt imfit}. Combining our measurements with those of \citet{Koss+2015ApJ807.149} at $8.4$ and $4.9$~GHz, we fit power laws of the form $S\propto\nu^\alpha$ to the spectral energy distributions, where $S$ is the integrated flux density, $\nu$ the frequency and $\alpha$ the spectral index, and estimate that the nuclear source has a spectral index $\alpha=-0.18\pm0.03$ while the south-west compact source (i.e.\ the approaching jet) has $\alpha=-0.8\pm0.3$ (see \autoref{fig:NGC3393_continuum}). These spectral indices are within the range of spectral indices of other nuclei and jets \citep[e.g.][]{Hovatta+2014AJ147.143}, and are consistent with self-absorbed optically thick synchrotron emission in the nucleus and optically thin synchrotron emission in the jet \citep[see e.g.][]{2019MNRAS.484.4239R, 2022MNRAS.510.4485R}. We note however that we have neither matched our resolution to that of the VLA observations nor applied a $15$~$\sigma$ cut as done by \citet{Koss+2015ApJ807.149}. These could lead to a bias in the spectral indices estimated.

\subsection{NGC~5765B} 
\label{sec:5765_obs}

\subsubsection{Molecular gas}

The \cotwo{} emission of NGC~5765B is very strong. The moment maps shown in \autoref{fig:NGC5765B_maps} show the typical morphology and dynamics of a barred disc galaxy, with emission particularly strong along two bisymmetric dust lanes (parallel to but offset from the optical bar; see the right column of \autoref{fig:optOver}) as well as two bisymmetric nuclear spiral arms.
The velocity field in turn shows a very strong kinematic twist beyond the dust lanes and thus the bar (most easily seen as a clear kink in the zero-velocity curve), and a milder kinematic twist in the centre within the nuclear spiral. We note that the sudden velocity jumps detected in the PVD within $\approx2\arcsec$ on either side of the galaxy centre are not the signatures of Keplerian rotation around a putative SMBH, but rather arise from the bar and the twisted velocity map. As expected, the velocity dispersion map shows broad line widths ($\sigma\gtrsim40$~km~s$^{-1}$) due to beam smearing along the offset dust lanes, that are probably tracing bisymmetric shocks along the leading edges of the bar \citep[see e.g.][] {Athanassoula+1992MNRAS259.345, 2006MNRAS.370.1499A, 2012ApJ...751..124K}, and in the very centre, that may also harbour gas with intrinsically high turbulence due to AGN and/or star-formation feedback. There are also mildly increased line widths slightly beyond the end of the dust lanes (slightly leading). The gas elsewhere is dynamically cold ($\sigma<20$~km~s$^{-1}$) and follows a regular rotation pattern. We do nevertheless note a slight excess of molecular gas in the northern half of the galaxy, that may be related to the gravitational interaction with NGC~5765B's companion galaxy NGC~5765A. Another galaxy with a very similar barred morphology has recently been studied at much higher spatial resolution ($24$~pc) as part of the WISDOM project (NGC~5806; \citealt{Choi_submitted}).

As shown in the bottom-left panel of \autoref{fig:NGC5765B_maps}, our JCMT \cotwo{} spectrum (integrated flux density $186\pm16$~Jy~km~s$^{-1}$, corresponding to an integrated molecular gas mass of $(6.0\pm0.5)\times10^{9}$~M$_\odot$) is in good agreement with our synthesised ALMA integrated spectrum ($216.2\pm0.4$~Jy~km~s$^{-1}$ and $(6.94\pm0.01)\times10^{9}$~M$_\odot$). The slight difference is likely due to the primary beam of the JCMT being slightly smaller than the total extent of the molecular gas detected with ALMA. Due to the relatively large distance of this galaxy, the ALMA primary beam reaches $\approx1.8$~$R_\mathrm{e}$, well beyond the total extent of the molecular gas detected, so all the molecular gas of NGC~5765B has probably been detected. The JCMT \cothree{} spectrum is very similar to the \cotwo{} spectrum, the small differences probably reflecting minor excitation and/or distribution differences.

\subsubsection{230-GHz continuum emission}

The $230$-GHz continuum map shown in the bottom-right panel of \autoref{fig:NGC5765B_maps} reveals one bright nuclear compact source (taking the VLBI source as the galaxy centre), a fainter compact source $\approx0\farcs4$ ($\approx200$~pc) to the south-west of the centre and fuzzy extended emission to the north-west of the centre. Considering their locations, compactness and relative orientation (perpendicular to the maser disc), the two compact sources are likely due to the AGN (the off-centre source perhaps tracing a young jet). The fuzzy emission to the north-west partially overlaps with high surface brightness and/or high velocity dispersion regions in the zeroth- and second-moment maps. It may be dust emission, but nothing is detected on the opposite side of the galaxy despite a bisymmetric molecular gas distribution. Multi-band radio observations should constrain the spectral indices and therefore help reveal the origin of these sources.

\section{Potential for SMBH Mass Measurements}
\label{sec_discussion}

Based on the detailed descriptions of the previous section, we now discuss the potential of each galaxy for a SMBH mass measurement using CO kinematics and future higher angular resolution ALMA observations ($\sim0\farcs01$).

NGC~1194 has a narrow, lopsided and disturbed molecular gas distribution (see \autoref{sec:1194_obs}) that would prevent any robust dynamical modelling. In any case, it has very little molecular gas (see \autoref{fig:NGC1194_maps}), and the faintness of the \cotwo{} emission would result in impractically long exposure times with any synthesised beam much smaller than the current one. For a quantitative estimate, we take the brightest pixel across the central synthesised beam and all channels of the current datacube (with a flux of $2.95$~mJy~beam$^{-1}$ at $4030$~km~s$^{-1}$), assume a homogeneous molecular gas distribution within it and adopt a smaller synthesised beam of $0\farcs045$ required to marginally spatially resolve the predicted angular radius of the SMBH SoI ($\theta_\mathrm{SoI}=0\farcs050$; see \autoref{tab:maserSOI}). Requiring a signal-to-noise ratio $S/N=5$ per synthesised beam and $10$~km~s$^{-1}$ channels, we estimate using the ALMA Observing Tool (OT) that the observations would require a total observing time of $\approx70$~days.

Another uncertainty of course is that while the gas kinematics is consistent with a regular rotating disc at the current angular resolution, it is possible that at higher angular resolutions the very centre (at the scale of the SMBH SoI) would be disturbed or have a much lower CO surface brightness (e.g.\ a central hole), as is the case in several galaxies \citep[e.g.][]{Davis+2018MNRAS473.3818, Smith+2019MNRAS485.4359, Ruffa_accepted}. This concern of course applies to all galaxies, those discussed here and others.

NGC~3393 has a fairly regular molecular gas distribution and kinematics at large spatial scales (see \autoref{sec:3393_obs}), but there are many sub-structures and kinematic disturbances at small scales, and \citet{Finlez+2018MNRAS479.3892} required both a large-scale bar and a nuclear bar to model the kinematics. Such kinematic complexity would make it extremely difficult to robustly infer a SMBH mass through dynamical modelling. This difficulty would be compounded by the faintness of the \cotwo{} emission in the very centre and the lack of gas along the kinematic major axis (see \autoref{fig:NGC3393_maps}), the regions that best constrain the SMBH mass. In any case, for a central peak intensity of $2.38$~mJy~beam$^{-1}$ at $3630$~km~s$^{-1}$, and requiring the smallest synthesised beam currently provided by ALMA ($0\farcs02$) to attempt to spatially resolve the predicted $R_\mathrm{SoI}$ ($\theta_\mathrm{SoI}=0\farcs013$; see \autoref{tab:maserSOI}), estimating the required ALMA observation time as above results in a total of $\approx100$~years.

The molecular gas distribution and kinematics of NGC~5765B are typical of those of barred disc galaxies (see \autoref{sec:5765_obs}). While this would make it challenging to model the large-scale kinematics, with sufficient spatial resolution it may be possible to model what appears to be a decoupled central disc (within the nuclear spiral and inner kinematic twist). The high velocity dispersions in the very centre are consistent with such a fast rotating disc. However, while having relatively bright \cotwo{} emission, with a peak central intensity of $27.4$~mJy~beam$^{-1}$ at $8190$~km~s$^{-1}$, the small angular scale required to resolve the predicted $R_\mathrm{SoI}$ ($\theta_\mathrm{SoI}=0\farcs013$; see \autoref{tab:maserSOI}) leads to an impossible integration time. Indeed, requiring a synthesised beam of $0\farcs02$ (the smallest currently available) and estimating the required ALMA observation time as above results in a total of $\approx80$~days.

Overall, because primarily of the impossibly long observation times required, and to a lesser extent the disturbed gas kinematics, none of our three target is ultimately suitable for a SMBH mass measurement using ALMA. Of course, we have assumed here that CO is the most abundant cold molecular gas tracer in these three galaxies, and that observing \cotwo{} yields the best balance between $S/N$ and angular resolution, but we cannot rule out the possibility that another cold molecular gas tracer (e.g.\ higher CO transition or higher-density tracer) might be better suited to measure the SMBH masses in these galaxies.

Apart from the intrinsic faintness of \cotwo{} in NGC~1194 and NGC~3393, the long observation times are primarily driven by the extremely small $\theta_\mathrm{SoI}$ required, as the observation time scales with the negative fourth power of $\theta_\mathrm{SoI}$ at a given surface brightness. In turn, the small $\theta_\mathrm{SoI}$ have primarily two causes. First, the maser method mostly probes low-mass SMBHs ($M_\mathrm{BH}\sim10^7$~M$_\odot$, lower than most successful SMBH mass measurements using cold molecular gas), yielding small $R_\mathrm{SoI}$. Second, because of the scarcity of masers, maser-hosting galaxies are on average rather distant, much farther than most galaxies with existing SMBH mass measurements (the few nearby potential targets considered in \autoref{sec:candidate} did not satisfy the other selection criteria), yielding small $\theta_\mathrm{SoI}$. To successfully cross-check the maser and cold molecular dynamics methods, maser-hosting galaxies that are both nearby and have regular dust/molecular gas distributions are required.

\section{Links between disc properties and maser emission}
\label{sec:maser_disturb}

The current ALMA observations can help uncover the relationship between the (central) molecular gas discs and masers. For example, all three galaxies studied here have a somewhat disturbed and/or clumpy molecular gas disc with a central mass concentration and likely non-circular motions.

To improve the number statistics and probe these trends further, we searched the literature for other published CO interferometric observations (i.e.\ moment maps) of galaxies in our parent sample of maser-hosting galaxies (\autoref{tab:maserSOI}). In addition to the aforementioned NGC~1386 and Circinus in \autoref{sec:candidate} (and the three galaxies presented in this paper), there are publications concerning the galaxies NGC~1068, NGC~2273, NGC~4388 and NGC~4945. All except NGC~2273 show features similar to those of the three galaxies presented in this paper. The \cotwo{} molecular gas in NGC~1386 has regular rotation on large scales (within $\approx1$~kpc in radius) but the centre ($\approx220$~pc in radius) is kinematically decoupled at a spatial resolution of $36$~pc, leading to significant residuals from axisymmetric models \citep{Ramakrishnan+2019MNRAS487.444}. Other kinematic kinks caused by the bar can also be seen in the \coone{} data presented by \citet{Zabel+2019MNRAS483.2251}. These features resemble those of NGC~5765B. The molecular gas in the Circinus Galaxy consists of nuclear spiral arms (within $\approx40$~pc in radius) and a circumnuclear disc ($\approx10$~pc in radius), the latter showing a highly distorted velocity field at a spatial resolution of $3$~pc \citep{2018ApJ...867...48I, 2022A&A...664A.142T}. NGC~1068 has a ring-shaped deficit of molecular gas ($\approx130$~pc outer diameter) surrounding a CO-rich nucleus, and it shows strong distortions (at a resolution of $6$~pc) in its velocity field both outside (i.e.\ beyond a radius of $\approx200$~pc) and within (i.e.\ within a radius of $\approx15$~pc) the ring-shaped deficit, presumably caused by AGN outflows \citep{2019A&A...632A..61G}. The \cotwo{} kinematics in the central $\approx500$~pc in radius of NGC~2273 is almost perfectly regular at a spatial resolution of $90\times72$~pc$^2$, although the \cotwo{} distribution is similar to that of NGC~5765B, showing evidence of a molecular gas-rich nuclear spiral \citep{2020A&A...643A.127D}.
NGC~4388 has a molecular gas depression in the central $\approx20$~pc in radius as well as prominent kpc-scale molecular gas outflows, the latter causing significant kinematic disturbances (at $12$-pc resolution) in the nucleus (i.e.\ within a radius of $\approx40$~pc; \citealt{2020A&A...643A.127D, 2021A&A...652A..98G}). In NGC~4945, prominent outflows and bar-driven inflows of molecular gas are traced by \cothree{} \citep{2021ApJ...923...83B} and multiple dense-gas tracers \citep{2018A&A...615A.155H}. The gas kinematics is highly disturbed within a radius of $\approx250$~pc at a resolution of $40$~pc (see e.g.\ Figure~8 of \citealt{2018A&A...615A.155H}), although a nuclear disc of $\approx50$~pc radius may be regularly rotating (see e.g.\ Figure~12 of \citealt{2018A&A...615A.155H}).

Of the nine maser galaxies with spatially resolved CO observations discussed above, almost all have morphological irregularities and/or kinematic disturbances and/or inflows/outflows, the only exception being NGC~2273 (that nevertheless shows potential bar-driven gas inflows). Although this sample is neither fully representative nor sufficiently large, the observations do suggest an emerging correlation between the properties of the central molecular gas disc and the existence of masers. It may be that a disturbed gas disc and/or gas inflows at kiloparsec scale is necessary to form a very dense molecular gas concentration at parsec scale, in turn triggering maser emission. Alternatively, it may be that AGN with masers are likely to cause irregularities in the gas discs, potentially through interactions between jets and the ISM.

As masers mostly reside in Seyfert~2 AGN, to understand whether this emerging trend is exclusive to maser galaxies or is more generally associated with the whole Seyfert~2 galaxy population, we searched the literature for other (non-maser-hosting) Seyfert~2 galaxies with published interferometric observations of cold molecular gas. \citet{2021A&A...653A.172S} reported a frequency of $53\%$ of outflows in a sample of $19$ AGN (mainly type~2), as part of the Physics at High Angular resolution in Nearby Galaxies (PHANGS) project. However, they did not discuss other non-circular motions. \citet{2020A&A...639A..43A} presented the zeroth-moment maps of $18$ Seyfert galaxies (including $10$ Seyfert~2) in their Figure~1. The sample was selected to have published mid-infrared spectral observations. All these galaxies have some morphological irregularities, such as non-axisymmetric gas distributions, off-centred peaks, holes/gaps and/or nuclear rings/spirals. This suggests that molecular gas irregularities are prevalent in the entire Seyfert~2 population. Nevertheless, the sample is still small in size, with unexplored potential biases, and critically the cold gas kinematics has not yet been explored.

Among publications with velocity maps available, all (non-maser-hosting) Seyfert~2 galaxies have features similar to those described in this work, e.g.\ Mrk~1066, NGC~7465 \citep{2020A&A...643A.127D}, NGC~4968 and NGC~4845 \citep{2021A&A...654A..24B}. However, the samples are even smaller in size and/or do not aim to be representative of all Seyfert~2 galaxies. Most publications concern a single object and aim to report non-circular motions in the first place. 

Therefore, it is difficult to draw any statistical conclusion about how prevalent irregular kinematic features are in (non-maser-hosting) Seyfert~2 galaxies, to contrast with the maser sample discussed in this section. In addition, the non-detection of masers in Seyfert~2 galaxies may well be due to inclination effects rather than the non-existence of masers. Thus, physical differences between maser-hosting and non-maser-hosting Seyfert~2 galaxies will be difficult to establish without large and carefully-constructed samples of both.

All in all, while the above discussion is inconclusive, further investigation of the trend reported is warranted and desirable.

\section{Summary and Conclusions}
\label{sec_conclusions}

Our primary goal was to identify galaxies with existing megamaser SMBH mass measurements that are also promising targets for future measurements using high angular resolution ($\sim0\farcs01$) ALMA molecular gas observations, to cross-check the two methods. Considering all galaxies with a megamaser SMBH mass measurement, three promising galaxies were identified (NGC~1194, NGC~3393 and NGC~5765B) and new ALMA intermediate angular resolution ($\approx0\farcs5$) and JCMT single-dish observations were obtained. The main results are as
follows.

\begin{enumerate}
\item NGC~1194 has an edge-on, disturbed and lopsided central \cotwo{} distribution dominated by two large components that appear associated with an irregular dust lane crossing the galaxy centre. The $230$-GHz continuum emission is dominated by a single compact nuclear source.

\item NGC~3393 has fairly regular but patchy \cotwo{} emission, with little gas near the kinematic major axis and only faint diffuse emission in the very centre. There are also two brighter structures in the central region that are reminiscent of (part of) a nuclear ring and/or spiral. The velocity field has kinematic twists typical of (doubly) barred disc galaxies. The $230$-GHz continuum emission is dominated by two compact sources. Combined with radio continuum flux densities from the literature, these reveal spectral indices typical of AGN/jets.
 
\item NGC~5765B has very bright \cotwo{} emission exhibiting the typical morphology and dynamics of a barred disc galaxy, with emission concentrated along two bisymmetric offset dust lanes (probably tracing shocks) and two bisymmetric nuclear spiral arms, with associated kinematic twists in the velocity field and large line widths probably due to the shocks. The $230$-GHz continuum emission is dominatedby a compact nuclear source and extended diffuse emission on one side of the nucleus.

\end{enumerate}

Overall, partially because of the disturbed molecular gas kinematics, but primarily because of the extremely long observation times required, none of the three galaxies is promising for a future SMBH mass measurement using molecular gas. These difficulties directly arise from the properties of maser-hosting galaxies: (i) frequent co-existence of masers and disturbed CO kinematics, as discussed in \autoref{sec:maser_disturb};
(ii) relatively low SMBH masses ($\sim10^7$~M$_\odot$), yielding small $R_\mathrm{SoI}$; and (iii) scarcity of masers, yielding  typically large galaxy distances and thus small $\theta_\mathrm{SoI}$.

Apart from our three target galaxies, other candidates could emerge if the parent sample of maser-hosting galaxies were enlarged and/or the selection criteria used in this paper were moderately relaxed.
In particular, by observing (different molecular lines) at higher frequencies, the criterion of resolving the putative SMBH SoI with a $0\farcs01$ synthesised beam could be relaxed (although the observing times required are likely to remain impractically long). In practice, the order-of-magnitude estimate of a SMBH SoI adopted in this paper (see \autoref{sec:candidate}) is also often smaller than actual measurements \citep[e.g.][]{2017MNRAS.466.1987Y}. 

Nonetheless, the new CO observations presented in this paper have significantly added to the rather small number of spatially-resolved molecular gas studies of maser-hosting galaxies. A detailed morphological and kinematical examination of our three targets, as well as six other maser-hosting galaxies with analogous observations from the literature, has revealed a potential correlation between molecular gas disturbances and/or inflows/outflows and the existence of maser emission.

\section*{Acknowledgements}
We thank the anonymous referee for the helpful comments. MDS acknowledges support from a Science and Technology Facilities Council (STFC) DPhil studentship ST/N504233/1. MB was supported by STFC consolidated grant `Astrophysics at Oxford' ST/K00106X/1 and ST/W000903/1. TAD acknowledges support from a STFC Ernest Rutherford Fellowship.
This paper used the following ALMA data: ADS/JAO.ALMA\#2015.1.00086.S and ADS/JAO.ALMA\#2016.1.01553.S. ALMA is a partnership of ESO (representing its member states), NSF (USA) and NINS (Japan), together with NRC (Canada), MOST and ASIAA (Taiwan), and KASI (Republic of Korea), in cooperation with the Republic of Chile. The Joint ALMA Observatory is operated by ESO, AUI/NRAO and NAOJ. The James Clerk Maxwell Telescope is operated by the East Asian Observatory on behalf of The National Astronomical Observatory of Japan, Academia Sinica Institute of Astronomy and Astrophysics, Korea Astronomy and Space Science Institute, National Astronomical Research Institute of Thailand, Center for Astronomical Mega-Science (as well as the National Key R\&D Program of China No.~2017YFA0402700). Additional funding support is provided by the Science and Technology Facilities Council of the United Kingdom and participating universities and organisations in the United Kingdom and Canada. The Starlink software \citep{2014ASPC..485..391C} is currently supported by the East Asian Observatory. This research used observations made with the NASA/ESA Hubble Space Telescope and obtained from the Hubble Legacy Archive, which is a collaboration between the Space Telescope Science Institute (STScI/NASA), the Space Telescope European Coordinating Facility (ST-ECF/ESA), and the Canadian Astronomy Data Centre (CADC/NRC/CSA). This research also used the NASA/IPAC Extragalactic Database (NED), which is operated by the Jet Propulsion Laboratory, California Institute of Technology, under contract with the National Aeronautics and Space Administration, NASA's Astrophysics Data System Bibliographic Services and Cube Analysis and Rendering Tool for Astronomy ({\tt CARTA}) for data visualisation and measurements \citep{2021ascl.soft03031C}.
The Digitized Sky Surveys were produced at the Space Telescope Science Institute under U.S. Government grant NAG W-2166.
Funding for the SDSS and SDSS-II has been provided by the Alfred P. Sloan Foundation, the Participating Institutions, the National Science Foundation, the U.S. Department of Energy, the National Aeronautics and Space Administration, the Japanese Monbukagakusho, the Max Planck Society, and the Higher Education Funding Council for England. The SDSS Web Site is \url{http://www.sdss.org/}.
The SDSS is managed by the Astrophysical Research Consortium for the Participating Institutions. The Participating Institutions are the American Museum of Natural History, Astrophysical Institute Potsdam, University of Basel, University of Cambridge, Case Western Reserve University, University of Chicago, Drexel University, Fermilab, the Institute for Advanced Study, the Japan Participation Group, Johns Hopkins University, the Joint Institute for Nuclear Astrophysics, the Kavli Institute for Particle Astrophysics and Cosmology, the Korean Scientist Group, the Chinese Academy of Sciences (LAMOST), Los Alamos National Laboratory, the Max-Planck-Institute for Astronomy (MPIA), the Max-Planck-Institute for Astrophysics (MPA), New Mexico State University, Ohio State University, University of Pittsburgh, University of Portsmouth, Princeton University, the United States Naval Observatory and the University of Washington.

\section*{Data Availability}
The raw ALMA data and the JCMT data obtained for this paper are available to download at the ALMA archive (\url{https://almascience.nrao.edu/asax/}) and the JCMT Science Archive (\url{https://www.eaobservatory.org/jcmt/science/archive/}), respectively. The {\it HST} images are available at the Hubble Legacy Archive (\url{https://hla.stsci.edu/hlaview.html#}). The final data products and original plots generated for the research underlying this article will be shared upon reasonable requests to the first author.

\bibliographystyle{mnras}
\bibliography{papers}


\bsp    
\label{lastpage}
\end{document}